\documentclass[prb,twocolumn,showpacs,floatfix,citesort]{revtex4}

\usepackage{graphicx}
\usepackage{amsmath}
\usepackage{amsfonts}
\usepackage{amssymb}

\begin{document}

\title{The Spin Density Matrix I: General Theory and Exact Master Equations}
\author{Sharif D. Kunikeev}
\affiliation{Department of Chemistry, University of Southern California, Los Angeles, CA
90089}
\author{Daniel A. Lidar}
\affiliation{Departments of Chemistry, Electrical Engineering, and Physics, University of
Southern California, Los Angeles, CA 90089}

\begin{abstract}
We consider a scenario where interacting electrons confined in quantum dots
(QDs) are either too close to be resolved, or we do not wish to apply
measurements that resolve them. Then the physical observable is an electron
spin only (one cannot unambiguously ascribe a spin to a QD) and the system
state is fully described by the spin-density matrix. Accounting for the
spatial degrees of freedom, we examine to what extent a Hamiltonian
description of the spin-only degrees of freedom is valid. We show that as
long as there is no coupling between singlet and triplet states this is
indeed the case, but when there is such a coupling there are open
systems effects, i.e., the dynamics is non-unitary even without
interaction with a true bath. Our primary focus is an investigation of non-unitary effects,
based on exact master equations we derive for the spin-density matrix in the
Lindblad and time-convolutionless (TCL) forms, and the implications for
quantum computation. In particular, we demonstrate that the Heisenberg
interaction does not affect the unitary part (apart from a Lamb shift) but
does affect the non-unitary contributions to time evolution of the
spin-density matrix. In a sequel paper we present a detailed analysis of an
example system of two quantum dots, including spin-orbit effects.
\end{abstract}

\maketitle

\section{Introduction}

In many quantum computation proposals the spin of a localized particle,
e.g., an electron or nuclear spin is a natural carrier of quantum
information -- a qubit -- which can be effectively protected and/or
processed to achieve a computational task. The ability to govern the spin
state via controllable interactions is a key ingredient underlying several
proposed scalable quantum computer architectures in semiconductor
nanostructures where the spin of an electron localized in a quantum dot (QD)
or by a donor atom serves as a single
qubit.\cite{Loss:98,Kane:98,Vrijen:00,HuSarma01,Schliemann01,HuSarma02,Koiller02,Kaplan04,Scarola,He05,Hu} 
It is a typical assumption that a single electron is trapped in each
individual QD and an electron spin can unambiguously be assigned to a QD.
Using the spin degree of freedom of electrons trapped in QDs (rather than
their charge) for information processing is of special interest since spins
have comparatively long coherence times in semiconductor nanostructures.\cite{KSS97,KA98,GAP98,AS02}

In this work we revisit a rather fundamental issue. In essence, we ask:
\textquotedblleft what is a spin?\textquotedblright . To clarify this
question (which is not meant in the sense of the relativistic origin of
spin, as our work is essentially nonrelativistic but includes relativistic
corrections such as spin-orbit interaction), we distinguish between the
notion of a \textit{pure} spin and a \textit{pseudo} spin.

Let us start with \textit{pseudo}-spins, a notion which applies to the
majority of studies utilizing spins for quantum information processing. In
order to ascribe a spin to a local site, such as a QD, one defines the
electron spin operator as a bilinear combination of electron annihilation
and creation Fermi operators, $c_{As}$, $c_{As}^{\dagger }$, in a localized
orbital $\phi _{A}$ ($s$ is a spin index, $A\ $is the QD\ index)%
\begin{equation}
s_{A}^{\alpha }=\frac{1}{2}\sum_{ss^{\prime }=1}^{2}c_{As}^{\dagger }\left(
\sigma _{\alpha }\right) _{ss^{\prime }}c_{As^{\prime }},\quad \alpha =x,y,z
\label{1.1}
\end{equation}%
(see, e.g., Appendix A in Ref.~\onlinecite{Aue}). Then the operators $\{s_{A}^{\alpha
}\}_{\alpha }$ obey the usual su$(2)$ commutation rules. Eq.~(\ref{1.1})
implies implicit dependence of the spin operators on coordinate degrees of
freedom and allows one to ascribe a spin to a QD. We call a spin defined by
Eq.~(\ref{1.1}) a \textit{pseudo-}spin, meaning that it carries some
coordinate dependence. It is important to note that the Heisenberg exchange
interaction arises only between \textit{pseudo}-spins. This is a consequence
of the coordinate dependence of pseudo-spins. Indeed, the Heisenberg
exchange interaction constant $J_{H}$ is electrostatic in nature, i.e., it
derives from the Coulomb interaction.\cite{Aue}

As opposed to a \textit{pseudo}-spin, we define a \textit{pure}-spin as a
spin that does not have any coordinate dependence. This is the usual
definition of a spin operator via the Pauli matrices $\vec{\sigma}=(\sigma
_{x},\sigma _{y},\sigma _{z})$,\cite{Landau} which do not depend on
coordinates. Considering the spin degrees of freedom as carriers of quantum
information, the spatial degrees of freedom must, in principle, be
irrelevant for the storage of quantum information. From the quantum
information point of view it does not matter in what orbital state an
electron is if one neglects possibly small spin-orbit interaction effects.
Although it is not possible, physically, to divide the system into purely
spin and spatial parts [as the \textquotedblleft spin\textquotedblright\
physics is embedded in space], we would like, in a succinct description of
spin dynamics, to eliminate irrelevant information. However, rather than
ignoring the coordinate dependence, we follow the standard procedure of open
quantum systems \cite{Breuer:book} and trace over the coordinates, leaving
us with a \emph{spin-density matrix }as the primary object of investigation.
Our motivation for investigating such a \emph{pure}-spin model is a scenario
where interacting and confined electrons either are too close to be
resolved, or one does not wish to apply measurements that resolve them. Then
the physical observable is an electron spin only (one cannot unambiguously
ascribe a spin to a site such as a QD) and the system state is fully
described by the spin-density matrix.

For simplicity of analysis we consider a prototype system of two interacting
electrons trapped in two sites $A$ and $B$ and separated by the distance $%
r_{AB}$. One of our first findings is
that the Heisenberg interaction, except for the Lamb energy shift, leaves the
unitary part of the pure-spin dynamics invariant. This means that the pure-spin
setting is inappropriate for any of the multitude of approaches to quantum
computation and decoherence control which rely on enacting quantum logic
gates via control of (sometimes only) Heisenberg
interactions.\cite{Loss:98,Kane:98,Burkard:99,Vrijen:00,Bacon:99a,Kempe:00,DiVincenzo:00a,Hu:01a,LidarWu:01,Bacon:Sydney,Bacon:01,Levy:01a,WuLidar:01b,Friesen:02,WuLidar:02a,Skinner:02,HuSarma01,Schliemann01,HuSarma02,Koiller02,MizelLidar:04,MizelLidar:04a,WoodworthMizelLidar:05,ByrdLidarWuZanardi:05,Kaplan04,Scarola,He05,Hu,WuLidarFriesen:04,FriesenBiswasHuLidar:07,Weinstein:05,Weinstein:07,Lidar:AQC-DD}
In these cases one must be able to \emph{resolve spins}, as in the 
pseudo-spin setting. After developing the appropriate formalism we turn to a
comparison of pure and pseudo-spin dynamics, in particular spin-orbit
effects.

In the case of \emph{pure}-spins we are particularly interested in finding
the conditions under which their dynamics is unitary (effectively closed
system)\ or non-unitary (open system). We show that as long as there is no
coupling between singlet and triplet states the dynamics is unitary (i.e., a
Hamiltonian description of the spin-only degrees of freedom is valid), but
with the singlet-triplet states coupling there are open systems effects,
i.e., the dynamics is non-unitary already due to the orbital degrees of
freedom, even without coupling to a true bath. To exhibit these effects we
derive several master equations for the spin-density matrix, and analyze
their implications for quantum computation. Our central results are
contained in Eqs. (\ref{Eq36}) and (\ref{tcl22}). Eq.~(\ref{Eq36}) is a
Lindblad-like master equation for the pure-spin dynamics, which clearly
exhibits the non-unitary nature of this dynamics (it does not,
however, invoke a Markovian approximation). Eq.~(\ref{tcl22}) does the
same within the time-convolutionless approach.

This paper is the first in a series of two. In this paper (part I), we
develop a variety of general models for pure-spin open system dynamics.
Specifically, in Section \ref{sec:OSR}, the operator sum representation
for the spin density matrix is derived, while the Lindblad-type and
time-convolutionless (TCL) master equations are considered respectively in
Sections \ref{sec:Lind} and \ref{sec:TCL}. We conclude Part I with a
discussion in Section \ref{sec:concI}.

In part II \cite{KLII} we highlight the differences and relationship between 
\emph{pseudo}\textit{-} and \emph{pure}-spin models, and provide a concrete
illustration in terms of a system of two quantum dots trapping one electron
each. In particular, we present calculation results demonstrating
non-unitary effects in the \emph{pure}-spin model due to both external
magnetic field inhomogeneity and spin-orbit interaction.

Atomic units, $\hbar =e=m_{e}=1$, $1/c\simeq 1/137$, are used throughout
the paper unless otherwise stated.

\section{The operator-sum representation}

\label{sec:OSR}

\subsection{Hamiltonian}

The total two-electron system Hamiltonian has the generic form%
\begin{equation}
\hat{H}=\hat{h}_{1}+\hat{h}_{2}+w_{12}  \label{Eq1}
\end{equation}%
Here\ $\hat{h}_{i}$, $i=1,2$ is a one-electron Pauli Hamiltonian which
includes spin-dependent terms%
\begin{equation}
\hat{h}_{i}=\frac{1}{2m^{\ast }}\left( \vec{p}_{i}+\frac{1}{c}\vec{A}\mathbf{%
(}\vec{r}_{i},t)\right) ^{2}+V_{\mathrm{Tr}}\mathbf{(}\vec{r}_{i},t)+\vec{B}%
\mathbf{(}\vec{r}_{i},t)\cdot \vec{s}_{i}  \label{Eq2}
\end{equation}%
where $m^{\ast }$ is the effective electron mass in the medium, $\mathbf{(}%
\vec{r}_{i},\vec{p}_{i})$ are the electrons' position and momentum
operators; $\vec{A}\mathbf{(}\vec{r}_{i},t)$ and $V_{\mathrm{Tr}}\mathbf{(}%
\vec{r}_{i},t)$ are, respectively, the vector potential and the trapping
potential, which has two minima at sites $\vec{r}_{A}$ and $\vec{r}_{B}$
where the electrons are localized. The magnetic fields%
\begin{equation}
\vec{B}\mathbf{(}\vec{r}_{i},t)=\vec{B}_{\mathrm{ex}}\mathbf{(}\vec{r}%
_{i},t)+\vec{B}_{\mathrm{so}}\mathbf{(}\vec{r}_{i},\vec{p}_{i},t)
\label{Eq3}
\end{equation}%
are due to the external (possibly spatially inhomogeneous) magnetic field $%
\vec{B}_{\mathrm{ex}}$ and the spin-orbit interaction field $\vec{B}_{%
\mathrm{so}}$ which can usually considered to be a small perturbation; $\vec{%
s}_{i}=\frac{1}{2}\vec{\sigma}_{i}$ is the spin vector of Pauli matrices
(here we included the gyromagnetic factor $g_{e}$ and Bohr magneton $\mu
_{B} $ in the definition of the\ magnetic fields so that magnetic fields are
measured in energy units). The two-electron interaction term%
\begin{equation}
w_{12}=V_{\mathrm{ee}}(r_{12})+V_{\mathrm{dip}}(\vec{s}_{1},\vec{s}_{2},\vec{%
r}_{12})  \label{Eq4}
\end{equation}%
consists of the interelectron electrostatic interaction potential $V_{%
\mathrm{ee}}(r_{12})=1/(\varepsilon r_{12})$, where $\varepsilon $ is the
dielectric constant of the medium, with $\varepsilon =1$ in vacuum, and the
spin-spin magnetic dipole interaction \cite{Messiah}
\begin{eqnarray}
V_{\mathrm{dip}} &=& 1.45\,\mathrm{meV}\left( \frac{\vec{s}_{1}\cdot \vec{s%
}_{2}\,r_{12}^{2}-3(\vec{s}_{1}\cdot \vec{r}_{12})(\vec{s}_{2}\cdot \vec{r}%
_{12})}{r_{12}^{5}} \right. \notag \\
&-& \left. \frac{8\pi }{3}\vec{s}_{1}\cdot \vec{s}_{2}\,\delta (\vec{r}%
_{12})\right)%
\label{Eq5}
\end{eqnarray}
which contains a term inversely proportional to the cube of interelectron
distance $r_{12}$, and the contact term proportional to the $\delta
$-function (coordinates and spins in Eq.~(\ref{Eq5}) are measured in
atomic units).

\subsection{Basis States}

If the system described by the Hamiltonian Eq.~(\ref{Eq1}) can be considered
as a closed system that does not interact with its environment, then
dynamics of the state $\Psi _{\mathrm{tot}}$ is governed fully by the
corresponding Schr\"{o}dinger equation. The state $\Psi _{\mathrm{tot}}(\vec{%
r},\sigma ,t)$ depends both on the electrons' spatial coordinates $\vec{r}%
\equiv (\vec{r}_{1},\vec{r}_{2})$ and the spin variables $\sigma \equiv
(\sigma _{1},\sigma _{2})$. Within the ground state approximation, which
consists of neglecting excited states, we assume that an electron can be
trapped in two ground orbital states $\phi _{A}$ and $\phi _{B}$ localized
near the sites $\vec{r}_{A}$ and $\vec{r}_{B}$ with energies $\varepsilon
_{A}$ and $\varepsilon _{B}$ respectively; $\phi _{A}$ and $\phi _{B}$ are
orthonormal states obtained in the same trapping potential. We do not assume 
\textit{a priori} any symmetry of the trapping potential. With two up and
down spin states $\chi _{\uparrow ,\downarrow }(\sigma )$ ($s_{z}\chi
_{\uparrow ,\downarrow }=\pm \frac{1}{2}\chi _{\uparrow ,\downarrow }$), the
one-electron basis comprises 4 states: $\phi _{A}(\vec{r}_{i})\chi
_{\uparrow ,\downarrow }(\sigma _{i})$ and $\phi _{B}(\vec{r}_{i})\chi
_{\uparrow ,\downarrow }(\sigma _{i})$. The corresponding two-electron
basis set is defined by Slater determinants.\cite{Landau}

The singlet subspace:

\begin{equation}
\Phi _{si}=f_{si}(\vec{r}\mathbf{)}\chi _{s}(\sigma ),\quad i=1,2,3
\label{Eq6}
\end{equation}%
where two-electron symmetric orbitals%
\begin{eqnarray*}
f_{s1}(\vec{r}\mathbf{)} &\mathbf{=}&\frac{1}{\sqrt{2}}\left( \phi _{A}(\vec{%
r}_{1})\phi _{B}(\vec{r}_{2})+\phi _{A}(\vec{r}_{2})\phi _{B}(\vec{r}%
_{1})\right) \\
f_{s2}(\vec{r}\mathbf{)} &\mathbf{=}&\phi _{A}(\vec{r}_{1})\phi _{A}(\vec{r}%
_{2}),\quad f_{s3}(\vec{r})\mathbf{=}\phi _{B}(\vec{r}_{1})\phi _{B}(\vec{r}%
_{2})
\end{eqnarray*}%
represent the states of single and double occupancies respectively,\ and the
antisymmetric singlet spin function%
\begin{equation*}
\chi _{s}=\frac{1}{\sqrt{2}}\left( \chi _{\uparrow }(\sigma _{1})\chi
_{\downarrow }(\sigma _{2})-\chi _{\uparrow }(\sigma _{2})\chi _{\downarrow
}(\sigma _{1})\right)
\end{equation*}%
describes the spin state with total spin $S=0$ and magnetic spin projection
number $M_{S}=0$;

The triplet subspace:%
\begin{equation}
\Phi _{ti}=f_{t}(\vec{r}\mathbf{)}\chi _{ti}(\sigma ),\quad i=1,2,3
\label{Eq7}
\end{equation}%
where a two-electron antisymmetric orbital%
\begin{equation*}
f_{t}(\vec{r}\mathbf{)=}\frac{1}{\sqrt{2}}\left( \phi _{A}(\vec{r}_{1})\phi
_{B}(\vec{r}_{2})-\phi _{A}(\vec{r}_{2})\phi _{B}(\vec{r}_{1})\right)
\end{equation*}%
and the triplet symmetric spin states%
\begin{eqnarray*}
\chi _{t1} &=&\chi _{\uparrow }(\sigma _{1})\chi _{\uparrow }(\sigma _{2}),
\\
\chi _{t2} &=&\frac{1}{\sqrt{2}}\left( \chi _{\uparrow }(\sigma _{1})\chi
_{\downarrow }(\sigma _{2})+\chi _{\uparrow }(\sigma _{2})\chi _{\downarrow
}(\sigma _{1})\right) , \\
\chi _{t3} &=&\chi _{\downarrow }(\sigma _{1})\chi _{\downarrow }(\sigma
_{2})
\end{eqnarray*}%
describe spin states with $S=1$ and $M_{S}=1,0,-1$ respectively.

Note that the six basis functions $\Phi _{si}$ and $\Phi _{ti}$ in Eqs. (\ref%
{Eq6}) and (\ref{Eq7}) are orthonormal, as are their orbital and spin
functions:%
\begin{eqnarray}
\left\langle f_{si}\right. \left\vert f_{sj}\right\rangle &=&\delta
_{ij},\quad \left\langle f_{si}\right. \left\vert f_{t}\right\rangle =0,
\label{Eq8} \\
\left\langle \chi _{ti}\right. \left\vert \chi _{tj}\right\rangle &=&\delta
_{ij},\quad \left\langle \chi _{ti}\right. \left\vert \chi _{s}\right\rangle
=0.  \notag
\end{eqnarray}

In the above basis set, we have%
\begin{equation}
\Psi _{\mathrm{tot}}(t)=\sum_{i=1}^{3}\left( a_{si}(t)\Phi
_{si}+a_{ti}(t)\Phi _{ti}\right)  \label{Eq9}
\end{equation}%
where the expansion coefficients are solutions of the Schr\"{o}dinger
equation%
\begin{eqnarray}
i\left\vert \dot{a}_{s}\right\rangle &=&H^{ss}\left\vert a_{s}\right\rangle
+H^{st}\left\vert a_{t}\right\rangle ,  \label{Eq10} \\
i\left\vert \dot{a}_{t}\right\rangle &=&H^{ts}\left\vert a_{s}\right\rangle
+H^{tt}\left\vert a_{t}\right\rangle  \notag
\end{eqnarray}%
where the column vectors $|a_{s}\rangle =(a_{s1},a_{s2},a_{s3})^{T}$ and $%
|a_{t}\rangle =(a_{t1},a_{t2},a_{t3})^{T}$ are vectors of singlet and
triplet amplitudes, and%
\begin{eqnarray}
H_{ij}^{ss} &=&\left\langle \Phi _{si}\right\vert \hat{H}\left\vert \Phi
_{sj}\right\rangle ,\quad H_{ij}^{st}=\left\langle \Phi _{si}\right\vert 
\hat{H}\left\vert \Phi _{tj}\right\rangle ,  \notag \\
H_{ij}^{ts} &=&\left\langle \Phi _{ti}\right\vert \hat{H}\left\vert \Phi
_{sj}\right\rangle ,\quad H_{ij}^{tt}=\left\langle \Phi _{ti}\right\vert 
\hat{H}\left\vert \Phi _{tj}\right\rangle ,  \label{Eq11} \\
H^{ss\dagger } &=&H^{ss},\quad H^{tt\dagger }=H^{tt},\quad H^{st\dagger
}=H^{ts}  \notag
\end{eqnarray}%
are correspondingly singlet-singlet, singlet-triplet, triplet-singlet, and
triplet-triplet subspace interaction Hamiltonians. We note that within the
Heitler-London (HL) approximation,\cite{Heitler} one neglects the double
occupancy states contribution in the expansion Eq.~(\ref{Eq9}).

\subsection{Spin Density Matrix}

By analogy with the theory of open quantum systems, one can formally
consider the spin subsystem as a \textquotedblleft system\textquotedblright\
while the orbital degrees of freedom belong to the \textquotedblleft
bath\textquotedblright . Then the total Hamiltonian Eq.~(\ref{Eq1}) governs
the evolution of this ``system + bath'', with the ``system'' Hamiltonian being
the Zeeman interaction term with a space-independent magnetic field (this
term may be absent), the spin-independent terms describing the bath, and the
spin-dependent terms (which depend both on spin and coordinate variables)
describing the interaction between system and bath.

A description of the open system in terms of a completely positive (CP) map
will result if (but not only if \cite{Rodriguez:07}) we assume that system
(S) and bath (B) are initially decoupled, so that the total initial density
matrix $\rho _{\mathrm{tot}}(t=0)=\left\vert \Psi _{\mathrm{tot}%
}(0)\right\rangle \left\langle \Psi _{\mathrm{tot}}(0)\right\vert $ is a
tensor product of the system and bath density matrices [$\rho (0)$ and $\rho
_{B}(0)$ respectively], with $\rho (0)$ being defined on the whole system
subspace. The system dynamics is described by the \emph{spin density matrix}:%
\begin{equation}
\rho (0)\longmapsto \rho (t)=\mathrm{Tr}_{B}\,\left[ U(t)(\rho (0)\otimes
\rho _{B}(0))U^{\dagger }(t)\right] .  \label{Eq12}
\end{equation}%
Here $\mathrm{Tr}_{B}$ is the partial trace over the coordinates
(\textquotedblleft bath\textquotedblright ) and the time-evolution operator
is%
\begin{equation}
U(t)=T_{\leftarrow }\,\exp \left( -i\int_{0}^{t}d\tau \,\hat{H}(\tau )\right) ,
\label{Eq13}
\end{equation}%
where $\hat{T}_{\leftarrow }$ is a chronological time-ordering operator.

Now we come to an important observation:\ it follows from Eqs.~(\ref{Eq6})
and (\ref{Eq7})\ that the spin and orbital degrees of freedom in $\left\vert
\Psi _{\mathrm{tot}}(0)\right\rangle $ [Eq.~(\ref{Eq9})] are factorized only
if the state belongs to either the singlet or the triplet subspace, but not
to a superposition of both. Therefore, the total initial density matrix
cannot be represented in a tensor product form if the initial state contains
both singlet and triplet parts. In order to understand the case of a CP map
description, we next consider these two cases.

\subsubsection{Singlet initial state}

In this case $\left\vert a_{t}(0)\right\rangle =0$ and the total density
matrix state takes the form%
\begin{equation*}
\rho _{\mathrm{tot}}(0)=\rho (0)\otimes \sum_{ij}a_{si}(0)a_{sj}^{\ast
}(0)\left\vert f_{si}\right\rangle \left\langle f_{sj}\right\vert
\end{equation*}%
where $\rho (0)=\mathbf{S}$ with $\mathbf{S}=\left\vert \chi
_{s}\right\rangle \left\langle \chi _{s}\right\vert $ being the projection
operator on the singlet subspace, and $\left\langle a_{s}(0)\right\vert
\left. a_{s}(0)\right\rangle =1$. With this initial state, Eq.~(\ref{Eq12})
can be rewritten similarly to a CP\ map in the operator sum representation as:\cite{Kraus,K83,Breuer:book}
\begin{equation}
\rho (t)=\sum_{i=1}^{3}\mathbf{A}_{si}(t)\rho (0)\mathbf{A}_{si}^{\dagger
}(t)+\mathbf{A}_{t}(t)\rho (0)\mathbf{A}_{t}^{\dagger }(t),  \label{Eq14}
\end{equation}%
where the Kraus operators $\left\{ \mathbf{A}_{si},\mathbf{A}_{t}\right\} $
expressed in terms of the $a_{s,t}(t)$ coefficients are%
\begin{equation}
\mathbf{A}_{si}(t)=a_{si}(t)\,\mathbf{S},\quad \mathbf{A}_{t}(t)=%
\sum_{i=1}^{3}a_{ti}(t)\,\mathbf{K}_{i}^{\dagger }  \label{Eq15}
\end{equation}%
with 
\begin{equation}
\mathbf{K}_{i}=\left\vert \chi _{s}\right\rangle \left\langle \chi
_{ti}\right\vert
\end{equation}%
being operators coupling the triplet and singlet subspaces. Note that the
Kraus operators depend on initial conditions via the dependence of $%
\left\vert a_{s}(t)\right\rangle $ and $\left\vert a_{t}(t)\right\rangle $
on the initial amplitudes $\left\vert a_{s}(0)\right\rangle $. This implies
that Eq.~(\ref{Eq14})\ does in fact \emph{not} represent a CP\ map -- more
on this below, in subsection \ref{analysis}. The first term in Eq.~(\ref%
{Eq14}), containing the sum over $i$, and the second one describe,
respectively, the singlet and triplet states contributions to $\rho (t)$.

Using the normalization condition for the $a$'s -- $\left\langle
a_{s}(t)\right\vert \left. a_{s}(t)\right\rangle +\left\langle
a_{t}(t)\right\vert \left. a_{t}(t)\right\rangle =1$ -- one derives the
normalization condition%
\begin{equation}
\sum_{i=1}^{3}\mathbf{A}_{si}^{\dagger }\mathbf{A}_{si}+\mathbf{A}%
_{t}^{\dagger }\mathbf{A}_{t}=\mathbf{S,}  \label{Eq16}
\end{equation}%
which guarantees preservation of the trace of $\rho $ in the case of a
singlet initial state.

\subsubsection{Triplet initial state}

Similarly, in the triplet case the initial state is specified by%
\begin{equation*}
\rho _{\mathrm{tot}}(0)=\rho (0)\otimes \left\vert f_{t}\right\rangle
\left\langle f_{t}\right\vert
\end{equation*}%
where 
\begin{equation}
\rho (0)=\sum_{ij}a_{ti}(0)a_{tj}^{\ast }(0)\mathbf{T}_{ij},
\end{equation}%
with 
\begin{equation}
\mathbf{T}_{ij}=\left\vert \chi _{ti}\right\rangle \left\langle \chi
_{tj}\right\vert
\end{equation}%
being coupling operators between triplet states, and $\left\langle
a_{t}(0)\right\vert \left. a_{t}(0)\right\rangle =1$. Then, the operator sum representation is
exactly the same as Eq.~(\ref{Eq14}) but the Kraus operators $\left\{ 
\mathbf{A}_{si},\mathbf{A}_{t}\right\} $ are defined differently as%
\begin{equation}
\mathbf{A}_{si}(t)=\sum_{j=1}^{3}U_{ij}^{st}(t)\,\mathbf{K}_{j},\quad 
\mathbf{A}_{t}(t)=\sum_{i,j=1}^{3}U_{ij}^{tt}(t)\,\mathbf{T}_{ij},
\label{Eq17}
\end{equation}%
where the evolution operator matrix elements are defined as solutions of the
differential matrix equation%
\begin{eqnarray}
&&\left\{ 
\begin{array}{c}
i\dot{U}=HU, \\ 
U(0)=I%
\end{array}%
\right.  \label{Eq18} \\
U &=&\left( 
\begin{array}{cc}
U^{ss} & U^{st} \\ 
U^{ts} & U^{tt}%
\end{array}%
\right) ,\quad H=\left( 
\begin{array}{cc}
H^{ss} & H^{st} \\ 
H^{ts} & H^{tt}%
\end{array}%
\right) .  \notag
\end{eqnarray}%
Here $I$ is the $6\times 6$ identity matrix. Notice the difference between the
singlet and triplet cases in the definition of the Kraus operators. Namely,
the triplet Kraus operators do not depend on the initial conditions, unlike
the singlet case. The initial triplet amplitude dependence is present only
in $\rho (0)$. Nevertheless, even in the triplet case we do not obtain a CP\
map, since there is a dependence of the Kraus operators on the domain (in
this case on the subspace of triplet states).

The normalization condition changes to%
\begin{equation}
\sum_{i=1}^{3}\mathbf{A}_{si}^{\dagger }\mathbf{A}_{si}+\mathbf{A}%
_{t}^{\dagger }\mathbf{A}_{t}=\mathbf{T}  \label{Eq19}
\end{equation}%
where $\mathbf{T}=\mathrm{Tr\,}\mathbf{T}_{ij}=\sum_{i}\mathbf{T}_{ii}$ is
the projection operator on triplet subspace.

\subsubsection{Mixed initial state}

Although a mixed initial state, which contains both singlet and triplet
parts, cannot be represented in a product form, a slight modification of
Kraus operators in Eq.~(\ref{Eq15}) according to%
\begin{equation}
\begin{array}{ccc}
\mathbf{A}_{si}^{(m)}(t) & = & \mathbf{A}_{si}(t)/\left\langle
a_{s}(0)\right\vert \left. a_{s}(0)\right\rangle ^{1/2}, \\ 
\mathbf{A}_{t}^{(m)}(t) & = & \mathbf{A}_{t}(t)/\left\langle
a_{s}(0)\right\vert \left. a_{s}(0)\right\rangle ^{1/2}%
\end{array}
\label{Eq19+1}
\end{equation}%
will provide the operator sum representation Eq.~(\ref{Eq14}) for the mixed initial state. The
normalization condition Eq.~(\ref{Eq16}) is accordingly renormalized so that
the right-hand side is divided by $\left\langle a_{s}(0)\right\vert \left.
a_{s}(0)\right\rangle $. If $\left\langle a_{s}(0)\right\vert \left.
a_{s}(0)\right\rangle =1$, the mixed case formulas go over into the singlet
ones.

\subsection{Analysis}

\label{analysis}

Note that in spite of having assumed a factorized initial state in the
singlet and triplet cases, the operator sum representation Eq.~(\ref{Eq14}), is \emph{not}
completely positive.\cite{Breuer:book} This is because the singlet and
triplet Kraus operators, Eqs. (\ref{Eq15}) and (\ref{Eq17}), depend upon
which initial state (singlet or triplet) is chosen. This happens since
in the general case, when the initial spin density matrix $\rho (0)$
contains both singlet and triplet parts, it is not possible physically to
realize such a state in the product form $\rho (0)\otimes \left\vert
f\right\rangle $ $\left\langle f\right\vert $, with $\left\vert
f\right\rangle $ being a reference coordinate wavefunction, in the total
Hilbert space. In this case, $\left\vert f\right\rangle $ would be
simultaneously symmetric and antisymmetric with respect to permutations of
coordinates, which can be realized only if $f\equiv 0$. Therefore, it not
possible to obtain Kraus operators independent of $\rho (0)$ in the general
mixed case. The Kraus operators in Eq.~(\ref{Eq19+1}) do depend on
$\rho (0)$. There is no contradiction to the Kraus representation
theorem,\cite{Kraus,K83,Breuer:book} which states that a map $\rho (0)\rightarrow \rho (t)$ has the Kraus
representation if and only if it is linear, completely positive, and trace
preserving, because there is a subtle difference between the Kraus
representation of a map and the Kraus representation of a state under the
action of a map. The Kraus operators describing the Kraus representation of
a map are independent of the state $\rho (0)$ while the Kraus operators
describing the Kraus representation of a state $\rho (0)$ under the action
of a map may be dependent on the initial\ state.\cite{Tong:04}

\section{Lindblad-type master equation}

\label{sec:Lind}

While the operator sum representation, Eq.~(\ref{Eq14}), can in principle be used to investigate
spin dynamics, it is not clear how to separate out the unitary evolution of
the (spin) system from the possibly non-unitary one, which results from the
system-bath coupling. The reason is that\ in general, each Kraus operator
will contain a contribution from both the unitary and the non-unitary
components of the evolution. In Ref.~\onlinecite{Lidar}, a general procedure was
devised to derive from the operator sum representation a CP Markovian master equation, where
unitary and non-unitary terms could be identified. The Markovian part
of the derivation was based
on a coarse-graining procedure. Here, we use a similar formal
approach, based on
explicit expressions for the spin density matrix, but avoid the coarse
graining step since we are not interested in the Markovian limit.

Using Eq.~(\ref{Eq15}) for the Kraus
operators, the operator sum representation Eq.~(\ref{Eq14}) can be reduced to%
\begin{equation}
\rho (t)=\left\langle a_{s}(t)\right. \left\vert a_{s}(t)\right\rangle 
\mathbf{S+}\sum_{ij=1}^{3}\left( \left\vert a_{t}(t)\right\rangle
\left\langle a_{t}(t)\right\vert \right) _{ij}\mathbf{T}_{ij}.  \label{Eq20}
\end{equation}%
It is straightforward to check that in the general mixed case where both
triplet and singlet components are present in the initial state Eq.~(\ref%
{Eq20}) holds true. After some algebraic manipulation one can transform the
time derivative of Eq.~(\ref{Eq20}) into:%
\begin{equation}
\frac{\partial \rho }{\partial t}=-i\left[ \mathbf{H}^{tt},\rho \right] +%
\mathcal{L}_{ts}\left( a_{s}(t),a_{t}(t)\right)  \label{Eq21}
\end{equation}%
where%
\begin{equation}
\mathbf{H}^{tt}=\sum_{ij=1}^{3}H_{ij}^{tt}\,\mathbf{T}_{ij}  \label{Eq22}
\end{equation}%
and%
\begin{eqnarray}
\mathcal{L}_{ts} &=&i\left( \mathrm{Tr\,}(G)\mathbf{S-}\sum_{ij=1}^{3}G_{ij}%
\mathbf{T}_{ij}\right)  \label{Eq23} \\
G_{ij} &=&F_{ij}-F_{ij}^{\dagger },\quad F_{ij}=\left( H^{ts}\left\vert
a_{s}(t)\right\rangle \left\langle a_{t}(t)\right\vert \right) _{ij}  \notag
\end{eqnarray}%
In the derivation of Eq.~(\ref{Eq21}), we have used Eq.~(\ref{Eq10}).
Clearly, if $H^{ts}\equiv 0$ (no singlet-triplet coupling), then $\mathcal{L}%
_{ts}\equiv 0$ and Eq.~(\ref{Eq21}) describes unitary evolution.

Since the orthogonal projection operators $\mathbf{S}$ and $\mathbf{T}$
commute with $\rho $, Eq.~(\ref{Eq21}) is invariant under the
transformation%
\begin{equation}
\mathbf{H}^{tt}\longmapsto \mathbf{H}^{tt}+E_{s}(t)\mathbf{S}+E_{t}(t)%
\mathbf{T}  \label{Eq24}
\end{equation}%
where $E_{s}(t)$ and $E_{t}(t)$ are any functions of time. In the limit $%
H^{ts}\equiv 0$, $\mathbf{H}^{tt}$ represents an effective spin Hamiltonian.
The Hamiltonian Eq.~(\ref{Eq24}) is clearly Hermitian if $E_{s}(t)$ and $%
E_{t}(t)$\ are real functions.

A procedure to obtain the effective spin Hamiltonian was proposed in
Ref.~\onlinecite{MizelLidar:04,MizelLidar:04a,WoodworthMizelLidar:05},
based on a comparison 
of the two expectation values%
\begin{equation}
\left\langle \Psi _{\mathrm{tot}}\right\vert \hat{H}\left\vert \Psi _{%
\mathrm{tot}}\right\rangle =\mathrm{Tr\,}(\mathbf{H}_{\rm spin}\rho ).
\label{Eq25}
\end{equation}%
From the general form of the density matrix Eq.~(\ref{Eq20}) and the
relationship Eq.~(\ref{Eq25}) it then follows that $\mathbf{H}_{\rm spin}$ must
have the general representation%
\begin{equation}
\mathbf{H}_{\rm spin}=E_{s}(t)\mathbf{S+}\sum_{ij=1}^{3}E_{tij}(t)\,\mathbf{T}%
_{ij}  \label{Eq26}
\end{equation}%
where the functions $E_{s}(t),$ $E_{tij}(t)\,$\ satisfy the Hermiticity
conditions: $E_{s}(t)=E_{s}^{\ast }(t),$ $E_{tij}(t)\,=E_{tji}^{\ast }(t)\,$. When $H^{ts}\equiv 0$ one derives from Eq.~(\ref{Eq25}) 
\begin{equation}
\begin{array}{lll}
E_{s}(t) & = & \left\langle a_{s}(t)\right\vert H^{ss}\left\vert
a_{s}(t)\right\rangle /\left\langle a_{s}(t)\right. \left\vert
a_{s}(t)\right\rangle , \\ 
E_{tij}(t) & = & H_{ij}^{tt}.%
\end{array}%
\quad   \label{Eq26-1}
\end{equation}

Note that the Hamiltonian Eq.~(\ref{Eq24}) fits the general representation
form Eq.~(\ref{Eq26}). Also, it is clear that the Hamiltonian (\ref{Eq26}) is symmetric with respect to permutation of spin indices since the
basis operators $\mathbf{S}$ and $\mathbf{T}_{ij}$ are symmetric with respect
to spin permutations. In terms of single-spin operators $\vec{s}_{1}$ and $%
\vec{s}_{2}$ they take the form%
\begin{equation}
\begin{array}{lll}
\mathbf{S} & = & \frac{1}{4}I-\vec{s}_{1}\cdot \vec{s}_{2},\quad \mathbf{T}%
_{11}=\frac{1}{4}I+\frac{1}{2}S_{z}+s_{1z}s_{2z}, \\ 
\mathbf{T}_{22} & = & \frac{1}{4}I+s_{1x}s_{2x}+s_{1y}s_{2y}-s_{1z}s_{2z},
\\ 
\mathbf{T}_{33} & = & \frac{1}{4}I-\frac{1}{2}S_{z}+s_{1z}s_{2z}, \\ 
\mathbf{T}_{12} & = & \frac{1}{\sqrt{2}}\left[ \frac{1}{2}S_{+}+J_{s}\right]
,\quad \mathbf{T}_{23}=\frac{1}{\sqrt{2}}\left[ \frac{1}{2}S_{+}-J_{s}\right]
, \\ 
\mathbf{T}_{13} & = & s_{1x}s_{2x}-s_{1y}s_{2y}+i\left(
s_{1x}s_{2y}+s_{2x}s_{1y}\right) , \\ 
\mathbf{T} & \mathbf{=} & \frac{3}{4}I+\vec{s}_{1}\cdot \vec{s}_{2},\quad 
\mathbf{T}_{21}=\mathbf{T}_{12}^{\dagger }, \\ 
\mathbf{T}_{31} & = & \mathbf{T}_{13}^{\dagger },\quad \mathbf{T}_{32}=%
\mathbf{T}_{23}^{\dagger }%
\end{array}
\label{Eq27}
\end{equation}

where%
\begin{equation*}
\begin{array}{lll}
J_{s} & = & s_{1z}s_{2x}+s_{1x}s_{2z}+i\left(
s_{1z}s_{2y}+s_{1y}s_{2z}\right) , \\ 
S_{\pm } & = & S_{x}\pm iS_{y},\quad \vec{S}=\vec{s}_{1}+\vec{s}_{2},%
\end{array}%
\end{equation*}
and $I$ is the $4\times 4$ identity matrix. At the same time, the singlet-triplet
basis operators $\mathbf{K}_{i}$ are asymmetric%
\begin{eqnarray}
\mathbf{K}_{1} &=&-\frac{i}{2\sqrt{2}}\left\{ \left( \vec{J}_{as}\right)
_{x}-i\left( \vec{J}_{as}\right) _{y}\right\} ,\quad \mathbf{K}_{2}=\frac{i}{%
2}\left( \vec{J}_{as}\right) _{z},  \notag \\
\mathbf{K}_{3} &=&\frac{i}{2\sqrt{2}}\left\{ \left( \vec{J}_{as}\right)
_{x}+i\left( \vec{J}_{as}\right) _{y}\right\}  \label{Eq28}
\end{eqnarray}%
where $\vec{J}_{as}=\left[ \vec{s}_{2}-\vec{s}_{1}\times \vec{S}\right] $.
This symmetry property of the Hamiltonian and spin density matrix is quite
general since it is a consequence of the orthogonality of the coordinate
wave functions in the singlet and triplet subspaces [Eq.~(\ref{Eq8})].
Moreover, it remains valid beyond the two-state approximation used here:\
with inclusion of excited states, spatial orbitals lying in singlet and
triplet subspaces will still be orthogonal to each other.

We emphasize that the recipe Eq.~(\ref{Eq25}) implicitly assumes that the
spin dynamics is unitary. Besides, observe that Eq.~(\ref{Eq25}) is not
invariant under the transformation $\mathbf{H}_{\rm spin}\mapsto \mathbf{H}%
_{\rm spin}+E_{s}(t)\mathbf{S}+E_{t}(t)\mathbf{T}$. Using Eqs. (\ref{Eq27}), one
can rewrite%
\begin{equation}
E_{s}(t)\mathbf{S}+E_{t}(t)\mathbf{T=}\left( \frac{1}{4}E_{s}(t)+\frac{3}{4}%
E_{t}(t)\right) I+J_{H}(t)\vec{s}_{1}\cdot \vec{s}_{2},  \label{Eq28-1}
\end{equation}%
where 
\begin{equation}
J_{H}(t)=E_{t}(t)-E_{s}(t)
\end{equation}%
and the last term, $H_{\mathrm{ex}}$, is the familiar Heisenberg exchange
interaction. \emph{Invariance of Eq.~(\ref{Eq21}) under the transformation
Eq.~(\ref{Eq24}) means that the spin density matrix will not change under a
unitary transformation induced by this Hamiltonian transformation}. In
particular, it follows that the unitary transformation induced by the
Heisenberg exchange interaction%
\begin{equation}
U_{H}(t)=\exp \left( -i\lambda _{H}(t)\vec{s}_{1}\cdot \vec{s}_{2}\right)
\label{Eq28-2}
\end{equation}%
where $\lambda _{H}(t)=\int_{0}^{t}dt^{\prime }\,J_{H}(t^{\prime })$ does
not affect the state:%
\begin{equation}
\rho (0)\mapsto \rho (t)=U_{H}(t)\rho (0)U_{H}^{\dagger }(t)=\rho (0),
\label{Eq28-3}
\end{equation}%
where $\rho (0)$ can be taken in the general form $\rho (0)=\rho _{s}\mathbf{%
S}+\sum_{ij}\rho _{tij}\mathbf{T}_{ij}$, with $\rho _{s}$ and $\rho _{tij}$
being arbitrary parameters specifying $\rho (0)$. It is easy to verify Eq.~(%
\ref{Eq28-3}) directly using the identities $\mathbf{S}^{2}=\mathbf{S,}$ $%
\mathbf{T}^{2}=\mathbf{T,}$ $\mathbf{ST=TS}=\mathbf{0}$.

The conclusion that the Heisenberg interaction does not affect pure-spin
dynamics has important implications for schemes that rely on this
interaction in order to enact universal quantum computation and/or
decoherence
control.\cite{Loss:98,Kane:98,Burkard:99,Vrijen:00,Bacon:99a,Kempe:00,DiVincenzo:00a,Hu:01a,LidarWu:01,Bacon:Sydney,Bacon:01,Levy:01a,WuLidar:01b,Friesen:02,WuLidar:02a,Skinner:02,HuSarma01,Schliemann01,HuSarma02,Koiller02,MizelLidar:04,MizelLidar:04a,WoodworthMizelLidar:05,ByrdLidarWuZanardi:05,Kaplan04,Scarola,He05,Hu,WuLidarFriesen:04,FriesenBiswasHuLidar:07,Weinstein:05,Weinstein:07,Lidar:AQC-DD} 
The pure-spin approach cannot be applied in these cases, i.e., one must be
able to resolve spins in order for Heisenberg-based quantum computation to
work.

However, this conclusion does not mean that the Heisenberg interaction plays
no role in \emph{pure}-spin dynamics. In general, this dynamics is
non-unitary and as we shall see below the constant $J_{H}$, characteristic
of the magnitude of the Heisenberg interaction, appears in both the non-unitary
term and in the unitary \emph{Lamb}-shift energy, in the $\rho (t)$ dynamics
described by Eq.~(\ref{Eq36}).

In the general case, in order to obtain a correct effective spin Hamiltonian
one needs to analyze the exact Eq.~(\ref{Eq21}). $\mathcal{L}_{ts}$ is seen
to be a bilinear matrix function of the $a_{s}(t)$ and $a_{t}(t)$ amplitudes
entering as a linear combination of their cross-products, while $\rho (t)$
is a quadratic matrix function of the $a_{s}(t)$ and $a_{t}(t)$ amplitudes
entering in separate combinations. Our goal is to express $\mathcal{L}_{ts}$
as a linear matrix function of $\rho $: $\mathcal{L}_{ts}=\mathcal{L}_{ts}%
\left[ \rho (t)\right] $, so that the total initial state information
recorded in the $a_{s}(t)$ and $a_{t}(t)$ amplitudes is compressed into $%
\rho $. To do so we first consider the relationship between the amplitudes
at the initial time $t=0$ and $t$ using the time-evolution operator Eq.~(\ref%
{Eq18})%
\begin{equation}
\left\{ 
\begin{array}{c}
a_{s}(t)=U^{ss}(t)a_{s}(0)+U^{st}(t)a_{t}(0) \\ 
a_{t}(t)=U^{ts}(t)a_{s}(0)+U^{tt}(t)a_{t}(0)%
\end{array}%
\right. .  \label{Eq29}
\end{equation}

Evidently, in general it is not possible to establish a one-to-one
correspondence between the $a_{s}(t)$ and $a_{t}(t)$ amplitudes. There
should be some correlation between the amplitudes at $t=0$. Let
us assume, for example, that initial conditions are set up such that we have
a linear relation between the amplitudes, given by a correlation matrix $R_{m}(0)$:%
\begin{equation}
a_{s}(0)=R_{m}(0)a_{t}(0).  \label{Eq29m}
\end{equation}%
Then, from Eqs. (\ref{Eq29}) and (\ref{Eq29m}) we obtain%
\begin{equation}
a_{s}(t)=R_{m}(t)a_{t}(t),  \label{Eq30m}
\end{equation}%
where%
\begin{equation}
R_{m}(t)=[U^{ss}(t)R_{m}(0)+U^{st}(t)][U^{ts}(t)R_{m}(0)+U^{tt}(t)]^{-1}.
\label{Eq30m+1}
\end{equation}%
We will refer to this as the mixed case because the spin-density matrix will
have both singlet and triplet components if both $a_{s}(0)$ and $a_{t}(0)$
are non-zero.

Two other special cases where it is possible to establish a one-to-one
correspondence between the $a_{s}(t)$ and $a_{t}(t)$ amplitudes are: (i)
singlet initial state, $a_{t}(0)=0$ and (ii) triplet initial state, $%
a_{s}(0)=0$. In the singlet $(R_{s})$ and triplet $(R_{t})$ cases, we have
respectively%
\begin{equation}
a_{s}(t)=R_{s,t}(t)a_{t}(t),  \label{Eq30}
\end{equation}%
where%
\begin{equation}
R_{s,t}(t)=\left\{ 
\begin{array}{cc}
U^{ss}(t)\left( U^{ts}(t)\right) ^{-1} & \text{singlet case} \\ 
U^{st}(t)\left( U^{tt}(t)\right) ^{-1} & \text{triplet case}%
\end{array}%
\right.  \label{Eq31}
\end{equation}%
The triplet case is a particular case of the mixed one when $%
R_{m}(0)a_{t}(0)=0$. Assuming the existence of the inverse operators (below
we consider this issue in detail), the matrix $F_{\alpha }$, $\alpha \in
\{s,t,m\}$ in Eq.~(\ref{Eq23}) can be rewritten as%
\begin{equation}
F_{\alpha }=Q_{\alpha }(t)\left\vert a_{t}(t)\right\rangle \left\langle
a_{t}(t)\right\vert  \label{Eq32}
\end{equation}%
where $Q_{\alpha }(t)=H^{ts}R_{\alpha }(t)$. In order to separate unitary
from non-unitary evolution, we resolve $Q_{\alpha }$ into Hermitian and anti-Hermitian parts:
\begin{equation}
\begin{array}{c}
P_{\alpha }=Q_{\alpha }+Q_{\alpha }^{\dagger } \\ 
D_{\alpha }=Q_{\alpha }-Q_{\alpha }^{\dagger }%
\end{array}
\label{Eq33}
\end{equation}%
Then, using the identities 
\begin{equation}
\begin{array}{c}
\mathbf{T}_{ij}=\mathbf{K}_{i}^{\dagger }\mathbf{K}_{j},\quad \mathbf{K}_{i}%
\mathbf{K}_{j}^{\dagger }=\delta _{ij}\mathbf{S,} \\ 
\mathbf{K}_{i}\rho \mathbf{K}_{j}^{\dagger }=\left( \left\vert
a_{t}(t)\right\rangle \left\langle a_{t}(t)\right\vert \right) _{ij}\mathbf{%
S,}%
\end{array}
\label{Eq34}
\end{equation}%
we obtain%
\begin{eqnarray}
\mathcal{L}_{ts}^{(\alpha )} &=&-\frac{i}{2}\left[ \mathbf{P}_{\alpha },\rho %
\right] +\frac{1}{2}\sum_{ij}(iD_{\alpha })_{ji}\times  \notag \\
&&\left( 2\mathbf{K}_{i}\rho \mathbf{K}_{j}^{\dagger }-\rho \mathbf{K}%
_{j}^{\dagger }\mathbf{K}_{i}-\mathbf{K}_{j}^{\dagger }\mathbf{K}_{i}\rho
\right)  \label{Eq35}
\end{eqnarray}%
where $\mathbf{P}_{\alpha }=\sum_{ij}\left( P_{\alpha }\right) _{ij}\mathbf{T%
}_{ij}$. The first term in Eq.~(\ref{Eq35}) describes the effect the
singlet-triplet coupling has on the unitary part of the system (spin)
dynamics, and \textquotedblleft renormalizes\textquotedblright\ the system
Hamiltonian (an analog of the \textit{Lamb shift}), while the second one --
proportional to the matrix $D_{\alpha }$ -- is responsible for non-unitary
effects in the spin dynamics. Including this \textit{Lamb shift} into the
effective spin Hamiltonian $\mathbf{\tilde{H}}^{tt}=\sum_{ij}\tilde{H}%
_{ij}^{tt}\mathbf{T}_{ij}=\mathbf{H}^{tt}+\frac{1}{2}\mathbf{P}_{\alpha }$,
we rewrite Eq.~(\ref{Eq21}) as%
\begin{eqnarray}
\frac{\partial \rho (t)}{\partial t} &=&-i\left[ \mathbf{\tilde{H}}%
^{tt},\rho (t)\right] +\frac{1}{2}\sum_{ij}\chi _{ij}\times  \notag \\
&&\left( \left[ \mathbf{K}_{i},\rho (t)\mathbf{K}_{j}^{\dagger }\right] +%
\left[ \mathbf{K}_{i}\rho (t),\mathbf{K}_{j}^{\dagger }\right] \right)
\label{Eq36}
\end{eqnarray}
where $\chi _{ij}=(iD_{\alpha })_{ji}$.

Eq.~(\ref{Eq36}) is one of our central results. It is parametrized by the
time-dependent functions $\tilde{H}_{ij}^{tt}(t)$ and $\chi _{ij}(t)$ which
in turn can be expressed in terms of matrix elements of the Hamiltonian $H$
[Eq.~(\ref{Eq18})]. Since these functions depend on which state (singlet/triplet/mixed) is specified as the initial state, $\rho (0)$, we
further consider the first two cases separately.

We restrict ourselves to the case of a time-independent Hamiltonian. The
general case can in principle be reduced to the time-independent one by
dividing up the time interval into small subintervals and then approximating
the Hamiltonian by average ones over each time-subinterval.

First we consider the eigenvalue problem%
\begin{equation}
\left( 
\begin{array}{cc}
H^{ss} & H^{st} \\ 
H^{ts} & H^{tt}%
\end{array}%
\right) \left( 
\begin{array}{c}
\left\vert e_{sk}\right\rangle \\ 
\left\vert e_{tk}\right\rangle%
\end{array}%
\right) =\varepsilon _{k}\left( 
\begin{array}{c}
\left\vert e_{sk}\right\rangle \\ 
\left\vert e_{tk}\right\rangle%
\end{array}%
\right) ,\quad k=1,\cdots ,6.  \label{Eq37}
\end{equation}%
Using the closure relation, one obtains%
\begin{eqnarray}
\chi _{s,t}^{T} &=&i\left( Q_{s,t}-Q_{s,t}^{\dagger }\right) ,  \label{Eq38}
\\
P_{s,t} &=&Q_{s,t}+Q_{s,t}^{\dagger }  \notag
\end{eqnarray}%
where%
\begin{eqnarray}
&&%
\begin{array}{rr}
Q_{t}= & \left\{ \left( \sum_{k}\exp (-i\varepsilon _{k}t)H^{ts}\left\vert
e_{sk}\right\rangle \left\langle e_{tk}\right\vert \right) \times \right. \\ 
& \left. \left( \sum_{k}\exp (-i\varepsilon _{k}t)\left\vert
e_{tk}\right\rangle \left\langle e_{tk}\right\vert \right) ^{-1}\right\} ,%
\end{array}
\label{Eq39} \\
&&%
\begin{array}{rr}
Q_{s}= & \left\{ \left( \sum_{k}\exp (-i\varepsilon _{k}t)H^{ts}\left\vert
e_{sk}\right\rangle \left\langle e_{sk}\right\vert \right) \times \right. \\ 
& \left. \left( \sum_{k}\exp (-i\varepsilon _{k}t)\left\vert
e_{tk}\right\rangle \left\langle e_{sk}\right\vert \right) ^{-1}\right\} ,%
\end{array}
\label{Eq40}
\end{eqnarray}%
are exact expressions in terms of exact solutions of the eigenvalue
(eigenvector) problem Eq.~(\ref{Eq37}). In general, Eq.~(\ref{Eq37}) can be
solved numerically using standard Hermitian symmetric eigenvalue problem
routines (such as in the LAPACK library). Notice the oscillatory
behavior predicted by Eqs. (\ref{Eq39}) and (\ref{Eq40}).

Let us consider an approximate solution to the eigenvalue problem assuming
that $H^{st}$ is a small perturbation. To zeroth order ($H^{st}\equiv 0$), the eigenvalue problem Eq.~(\ref{Eq37}) is reduced to the separate
singlet and triplet subspace eigenvalue problems%
\begin{equation}
\begin{array}{c}
\left( H^{ss}-\varepsilon _{k}^{(0)}I\right) \left\vert
e_{sk}^{(0)}\right\rangle =0,\quad k=1,2,3, \\ 
\left( H^{tt}-\varepsilon _{k}^{(0)}I\right) \left\vert
e_{tk}^{(0)}\right\rangle =0,\quad k=4,5,6%
\end{array}
\label{Eq41}
\end{equation}%
and $\left\vert e_{sk}^{(0)}\right\rangle =\left\vert 0\right\rangle ,$ $%
k=4,5,6,$ $\left\vert e_{tk}^{(0)}\right\rangle =\left\vert 0\right\rangle ,$
$k=1,2,3$ where $\left\vert 0\right\rangle $ is a zero vector. Up to
first order in $H^{st}$, one obtains
\begin{equation}
\begin{array}{c}
\left\{ 
\begin{array}{l}
\left\vert e_{sk}^{(1)}\right\rangle =\left\vert e_{sk}^{(0)}\right\rangle ,
\\ 
\left\vert e_{tk}^{(1)}\right\rangle =\left( \varepsilon
_{k}^{(0)}I-H^{tt}\right) ^{-1}H^{ts}\left\vert e_{sk}^{(0)}\right\rangle ,%
\end{array}%
\right. \quad k=1,2,3, \\ 
\left\{ 
\begin{array}{l}
\left\vert e_{tk}^{(1)}\right\rangle =\left\vert e_{tk}^{(0)}\right\rangle ,
\\ 
\left\vert e_{sk}^{(1)}\right\rangle =\left( \varepsilon
_{k}^{(0)}I-H^{ss}\right) ^{-1}H^{st}\left\vert e_{tk}^{(0)}\right\rangle ,%
\end{array}%
\right. \quad k=4,5,6%
\end{array}
\label{Eq42}
\end{equation}

\subsection{Triplet initial state}

After substitution of the vectors Eq.~(\ref{Eq42}) into $Q_{t}$ Eq.~(\ref%
{Eq39}) we have%
\begin{eqnarray}
Q_{t} &=&\left( \sum\limits_{k=1}^{3}\sum\limits_{l=4}^{6}\frac{\exp
(-i\varepsilon _{k}^{(0)}t)-\exp (-i\varepsilon _{l}^{(0)}t)}{\varepsilon
_{k}^{(0)}-\varepsilon _{l}^{(0)}}H^{ts}\right. \times  \notag \\
&&\left. \left\vert e_{sk}^{(0)}\right\rangle \left\langle
e_{sk}^{(0)}\right\vert H^{st}\left\vert e_{tl}^{(0)}\right\rangle
\left\langle e_{tl}^{(0)}\right\vert \right) \times  \notag \\
&&\left( \sum\limits_{l=4}^{6}\exp (-i\varepsilon _{l}^{(0)}t)\left\vert
e_{tl}^{(0)}\right\rangle \left\langle e_{tl}^{(0)}\right\vert \right. +
\\
&&\sum\limits_{k=1}^{3}\sum\limits_{l,m=4}^{6}\exp (-i\varepsilon
_{k}^{(0)}t)\frac{\left\vert e_{tl}^{(0)}\right\rangle \left\langle
e_{tl}^{(0)}\right\vert H^{ts}\left\vert e_{sk}^{(0)}\right\rangle }{\left(
\varepsilon _{k}^{(0)}-\varepsilon _{l}^{(0)}\right) \left( \varepsilon
_{k}^{(0)}-\varepsilon _{m}^{(0)}\right) }\times  \notag \\
&&\left. \left\langle e_{sk}^{(0)}\right\vert H^{st}\left\vert
e_{tm}^{(0)}\right\rangle \left\langle e_{tm}^{(0)}\right\vert \right) ^{-1}
\label{Eq43} 
\end{eqnarray}

Neglecting the splitting of triplet energies, which is mainly due to the
small magnetic Zeeman interaction, one can replace $\varepsilon
_{k}^{(0)}\equiv \varepsilon _{t},$ $k=4,5,6;$ $\ \varepsilon _{t}=\frac{1}{3%
}\sum_{k=4}^{6}\varepsilon _{k}^{(0)}$. Also, due to the strong
interelectron repulsion in double occupancy states the difference of
energies in doubly occupied and triplet states is much larger than the
difference of energies in the singlet, singly occupied state and the triplet
ones (denote this energy difference, the Heisenberg exchange interaction
constant, by$\ J_{H}=\varepsilon _{t}-\varepsilon _{s})$ so that we can
safely neglect the contributions from the doubly occupied states. Within
these approximations, Eq.~(\ref{Eq43}) and (\ref{Eq38}) can be reduced to%
\begin{equation}
\begin{array}{ll}
Q_{t}= & \frac{1-\exp (iJ_{H}t)}{J_{H}}H^{ts}\left\vert
e_{s1}^{(0)}\right\rangle \left\langle e_{s1}^{(0)}\right\vert H^{st}, \\ 
\chi _{t}^{T}= & 2\frac{\sin (J_{H}t)}{J_{H}}H^{ts}\left\vert
e_{s1}^{(0)}\right\rangle \left\langle e_{s1}^{(0)}\right\vert H^{st}, \\ 
P_{t}= & 2\frac{1-\cos (J_{H}t)}{J_{H}}H^{ts}\left\vert
e_{s1}^{(0)}\right\rangle \left\langle e_{s1}^{(0)}\right\vert H^{st},%
\end{array}
\label{Eq44}
\end{equation}%
where $\left\vert e_{s1}^{(0)}\right\rangle $ is the singlet, singly
occupied state ($\left\vert e_{s2,3}^{(0)}\right\rangle $ are
correspondingly doubly occupied states).

Note that $\chi _{t}^{T}$ is not necessarily a positive definite matrix.
Indeed, since 
\begin{equation*}
H^{ts}\left\vert e_{s1}^{(0)}\right\rangle \left\langle
e_{s1}^{(0)}\right\vert H^{st}
\end{equation*}%
is a positive Hermitian matrix, the sign of the oscillatory function $%
g(t)=2\sin (J_{H}t)/J_{H}$ determines whether $\chi _{t}^{T}$ is positive
(when $g(t)>0$) or negative ($g(t)<0$). Additionally, this conclusion is
supported by the fact that one could in principle evolve the state in the
backward time direction. In this case, $g(t)<0$ even at small $t$.

Alternatively, and for consistency, one can use the direct asymptotic
expansions at small $t$ in order to derive Eqs. (\ref{Eq44}): 
\begin{equation}
\begin{array}{c}
U^{st}(t)=U^{st}(0)+\dot{U}^{st}(0)t+\ddot{U}^{st}(0)\frac{t^{2}}{2}+\cdots ,
\\ 
U^{tt}(t)=U^{tt}(0)+\dot{U}^{tt}(0)t+\ddot{U}^{tt}(0)\frac{t^{2}}{2}+\cdots .%
\end{array}
\label{Eq45}
\end{equation}%
Obtaining from Eq.~(\ref{Eq18}) explicit expressions for the time
derivatives, Eqs. (\ref{Eq45}) take the form%
\begin{equation}
\begin{array}{c}
U^{st}(t)=-iH^{st}t-\left( H^{ss}H^{st}+H^{st}H^{tt}\right) \frac{t^{2}}{2}%
+\cdots , \\ 
U^{tt}(t)=I-iH^{tt}t-\left( H^{ts}H^{st}+(H^{tt})^{2}\right) \frac{t^{2}}{2}%
+\cdots .%
\end{array}
\label{Eq46}
\end{equation}%
Inserting these asymptotic expressions into Eq.~(\ref{Eq38}), we obtain%
\begin{equation}
\begin{array}{c}
\begin{array}{ll}
\chi _{t}^{T}= & 2H^{ts}H^{st}t,%
\end{array}
\\ 
\begin{array}{ll}
P_{t}= & \left( H^{ts}H^{st}H^{tt}-H^{ts}H^{ss}H^{st}\right) t^{2}.%
\end{array}%
\end{array}
\label{Eq47}
\end{equation}%
It can be easily verified that the asymptotic expansions (\ref{Eq47})
coincide with the corresponding expressions (\ref{Eq44}) at small $t$
(lifting some approximations made above concerning singlet and triplet
energies).

Also, using Eq.~(\ref{Eq46}) one can calculate the triplet
states population at small $t$:
\begin{eqnarray}
\left\langle a_{t}(t)| a_{t}(t)\right\rangle & = & 
\left\langle a_{t}(0)\right\vert \left( U^{tt}(t)\right) ^{\dagger
}U^{tt}(t)\left\vert a_{t}(0)\right\rangle = 1-\alpha _{t}t^{2},
\notag \\ 
\alpha _{t} & = & \left\langle a_{t}(0)\right\vert H^{ts}\left(
H^{ts}\right) ^{\dagger }\left\vert a_{t}(0)\right\rangle%
\label{Eq48}
\end{eqnarray}
Observe that the triplet state probability decreases with
time, while the singlet state probability, $\left\langle a_{s}(t)\right.
\left\vert a_{s}(t)\right\rangle =1-\left\langle a_{t}(t)\right. \left\vert
a_{t}(t)\right\rangle ,$ increases due to the singlet-triplet coupling, with
the constant $\alpha _{t}$ being quadratically dependent on the
singlet-triplet interaction matrix $H^{ts}$. In other words, the total spin
of the system is not conserved if there is a non-zero coupling between
singlet and triplet states.

\subsection{Purity for a triplet initial state}

The time dependence of the purity $p(t)\equiv \mathrm{Tr\,}\rho ^{2}(t)$ may
serve as a measure of non-unitarity in dynamics [$p(t)=1$ iff the state is
pure; $p(t)$ is constant iff the dynamics is unitary]. With Eq.~(\ref{Eq48})
we have%
\begin{eqnarray}
p(t) &=&\mathrm{Tr\,}\rho ^{2}(t)=1-2\left\langle a_{t}(t)\right. \left\vert
a_{t}(t)\right\rangle +2\left\langle a_{t}(t)\right. \left\vert
a_{t}(t)\right\rangle ^{2}  \notag \\
&=&1-2\alpha _{t}t^{2}  \label{Eq49-1}
\end{eqnarray}%
at small $t$. Observe that non-unitary effects are proportional to the
constant $2\alpha _{t}$ which can also be defined as the expectation value
of the time derivative of the matrix $\chi _{t}^{T}$ in Eq.~(\ref{Eq47}): $%
2\alpha _{t}=\mathrm{Tr}[\mathrm{\,}\mathbf{\dot{\chi}}_{t}^{T}(t)\rho (0)]$%
, where $\mathbf{\dot{\chi}}_{t}^{T}=\sum_{ij}\dot{\chi}_{tij}^{T}(t)\mathbf{%
T}_{ij}$. In Part II \cite{KLII} we investigate numerically how the
triplet states population, purity, and \emph{Lamb}-shift depend on
time and other relevant physical parameters.

\subsection{Singlet initial state}

This case differs qualitatively from the triplet one by the fact that the
inverse operator in $Q_{s}$ Eq.~(\ref{Eq40}) is singular at $t=0$. From the asymptotic expansions%
\begin{equation}
\begin{array}{c}
U^{ts}(t)=-iH^{ts}t-\left( H^{ts}H^{ss}+H^{tt}H^{ts}\right) \frac{t^{2}}{2}%
+\cdots , \\ 
U^{ss}(t)=I-iH^{ss}t-\left( H^{st}H^{ts}+(H^{ss})^{2}\right) \frac{t^{2}}{2}%
+\cdots .%
\end{array}
\label{Eq50}
\end{equation}%
at small $t$ one finds%
\begin{equation}
\begin{array}{cc}
Q_{s}= & \frac{i}{t}H^{ts}\left( I-iH^{ss}t-\left(
H^{st}H^{ts}+(H^{ss})^{2}\right) \frac{t^{2}}{2}\right) \times \\ 
& \left( H^{ts}-i\left( H^{ts}H^{ss}+H^{tt}H^{ts}\right) \frac{t}{2}\right)
^{-1}%
\end{array}
\label{Eq51}
\end{equation}%
Assuming that the matrix $H^{ts}$ is non-singular, $Q_{s}=\frac{i}{t}I$ as $%
t\rightarrow 0$. However, since $Q_{s}$ is defined on the vectors $%
\left\vert a_{t}(t)\right\rangle $, it operates on the outer product $%
\left\vert a_{t}(t)\right\rangle \left\langle a_{t}(t)\right\vert $ in $%
F_{s} $ Eq.~(\ref{Eq32}) and $\left\vert a_{t}(t)\right\rangle
=-iH^{ts}t\left\vert a_{s}(0)\right\rangle $ at small $t$. Therefore, their
product, $F_{s}=i\,tH^{ts}\left\vert a_{s}(0)\right\rangle \left\langle
a_{s}(0)\right\vert H^{st}$, goes to zero as $t\rightarrow 0$. If $H^{ts}$
is a singular matrix and $\left\vert a_{s}(0)\right\rangle $ belongs
to the kernel of $H^{ts}$ then $Q_{s}=O(t^{-2})$ but $\left\vert
a_{t}(t)\right\rangle =-\left( H^{ts}H^{ss}+H^{tt}H^{ts}\right) \frac{t^{2}}{%
2}\left\vert a_{s}(0)\right\rangle $ and $F_{s}=O(t^{2})$ as
$t\rightarrow 0$. When $\left\vert a_{s}(0)\right\rangle $ \
does not belong to 
the kernel of $H^{ts}$, the \textit{pseudo}-inverse of $H^{ts}$ is
defined in the Moor-Penrose sense (see, e.g., Ref.~\onlinecite{GL96} and references
therein). Thus, in spite of the singular behavior of $Q_{s}$ at zero we have
regular behavior of all the corresponding terms in Eq.~(\ref{Eq36}) and the
operator $Q_{s}$ is well defined at $t\neq 0$ (it is not defined in the
pathological case where $H^{ts}\equiv 0$ but in this case there is no
connection between singlet and triplet subspaces and $\mathcal{L}_{ts}\equiv
0$).

Keeping up to the second order terms, Eq.~(\ref{Eq51}) can be rewritten as%
\begin{equation}
Q_{s}=\frac{i}{t}I-\frac{1}{2}\left( H^{tt}-\tilde{H}^{ss}\right) +\frac{it}{%
2}\left( \tilde{H}^{ss}-H^{ts}H^{st}\right)  \label{Eq52}
\end{equation}%
where $\tilde{H}^{ss}=H^{ts}H^{ss}\left( H^{ts}\right) ^{-1}$ is the
similarity transformed matrix $H^{ss}$ and one obtains 
\begin{equation}
\begin{array}{lll}
\chi _{s}^{T} & = & -\frac{2}{t}I+\frac{i}{2}\left( \tilde{H}^{ss}-\tilde{H}%
^{ss\dagger }\right) - \\ 
&  & \frac{t}{2}\left( \tilde{H}^{ss}+\tilde{H}^{ss\dagger
}-2H^{ts}H^{st}\right) , \\ 
P_{s} & = & -H^{tt}+\frac{1}{2}\left( \tilde{H}^{ss}+\tilde{H}^{ss\dagger
}\right) +\frac{it}{2}\left( \tilde{H}^{ss}-\tilde{H}^{ss\dagger }\right) .%
\end{array}
\label{Eq53}
\end{equation}

Comparing Eqs. (\ref{Eq47}) and (\ref{Eq53}), notice the qualitative
difference between singlet and triplet cases. While in the triplet case the
corresponding matrices are basically of the second order in the $H^{ts}$
interaction coupling, the singlet state matrices, $\chi _{s}^{T}$ and $P_{s}$%
, are of the order of $1/t$, $H^{tt}$, and $H^{ss}$, respectively, at small $%
t$. Observe that at small $t$, $\chi _{s}^{T}$ is positive when $t<0$ and
negative when $t>0$.

Similarly to the triplet case above, we compute the triplet states
population
\begin{eqnarray}
\left\langle a_{t}(t)| a_{t}(t)\right\rangle & = & 
\left\langle a_{s}(0)\right\vert \left( U^{ts}(t)\right) ^{\dagger
}U^{ts}(t)\left\vert a_{s}(0)\right\rangle = \alpha _{s}t^{2}, \notag \\ 
\alpha _{s} & = & \left\langle a_{s}(0)\right\vert H^{st}\left(
H^{st}\right) ^{\dagger }\left\vert a_{s}(0)\right\rangle .
\label{Eq56}
\end{eqnarray}
Here the constant $\alpha _{s}$ is the average of the $H^{st}\left(
H^{st}\right) ^{\dagger }$ interaction operator over the initial state $%
\left\vert a_{s}(0)\right\rangle $.

\subsection{Purity for a singlet initial state}

Now we have for the purity%
\begin{equation}
p(t)=\mathrm{Tr\,}\rho ^{2}(t)=1-2\alpha _{s}t^{2}.  \label{Eq57}
\end{equation}

Alternatively, the rate of non-unitarity, $2\alpha _{s}$, can be defined as
the expectation value of the time derivative of the matrix $\chi _{s}^{T}$,
now in the triplet states $\left\vert a_{t}(t)\right\rangle $, since $\chi
_{s}^{T}$ is defined on $\left\vert a_{t}(t)\right\rangle $ and $a_{t}(0)=0$:%
\begin{equation}
2\alpha _{s}=\left\langle a_{t}(t)\right\vert \dot{\chi}_{s}^{T}\left\vert
a_{t}(t)\right\rangle =\mathrm{Tr\,}[\mathbf{\dot{\chi}}_{s}^{T}(t)\rho (t)],
\label{Eq58}
\end{equation}%
where $\left\vert a_{t}(t)\right\rangle =-iH^{ts}t\left\vert
a_{s}(0)\right\rangle $ and $\dot{\chi}_{s}^{T}=(2/t^{2})I$, $\mathbf{\dot{%
\chi}}_{s}^{T}(t)=\sum_{ij}\dot{\chi}_{sij}^{T}(t)\mathbf{T}_{ij}$ at small $%
t$.

\section{The time-convolutionless form of the master equation}

\label{sec:TCL}

In the previous section we have examined the possibility of constructing a
master equation based on the explicit structure of the spin density matrix
in terms of the $a_{s,t}(t)$ amplitudes. We found that a meaningful 
\textit{quasi-closed} form of the master equation can be derived if the initial
state is assumed to be in a product form of orbital and spin functions,
correspondingly in singlet and triplet states, or in a correlated mixed
state. The \textit{quasi-closed} character of the equations we
obtained stems
from the fact that the matrix operators in Eq.~(\ref{Eq36}) are defined
differently in the singlet and triplet cases, although they do not depend on
the initial state amplitudes, $a_{s}(0)$ and $a_{t}(0)$ correspondingly. In
the correlated mixed case we have a correlation matrix $R_{m}(0)$
fixing the
relation between $a_{s}(0)$ and $a_{t}(0)$ amplitudes, and this
correlation information is present in the operators appearing in the
dynamical equation [cf. Eqs. (\ref{Eq29m})-(\ref{Eq36})].

In the present section we approach the problem of constructing the dynamical
equation using projection operator techniques, as applied in the derivation
of the TCL master equation.\cite{Breuer:book}

\subsection{Derivation of the TCL\ master equation}

For the analysis of the general mixed case we start with the exact equation
for $\rho _{\mathrm{tot}}$%
\begin{equation}
\dot{\rho}_{\mathrm{tot}}=-i\left[ H,\rho _{\mathrm{tot}}\right] =-iL\rho _{%
\mathrm{tot}}  \label{tcl1}
\end{equation}%
where $L=L^{\dagger }$ is the Liouvillean superoperator corresponding to $H$%
. The density matrix $\rho _{\mathrm{tot}}$ contains both relevant and
irrelevant information about the coordinate dependence (bath), which will be
averaged out after integration over the coordinates: $\rho =\mathrm{Tr}_{%
\vec{r}}\mathrm{\,}\rho _{\mathrm{tot}}$. Let us assume that the irrelevant
information can be eliminated by virtue of a time-independent projection
operator $P$, and the relevant information is assumed to be given by the
projected density matrix $P\rho _{\mathrm{tot}}$. Using the projection
operator technique,\cite{Shibata1,Shibata2,Breuer:book} one obtains the
well-known TCL master equation for $P\rho _{\mathrm{tot}}$%
\begin{eqnarray}
P\dot{\rho}_{\mathrm{tot}}(t) &=&K(t)P\rho _{\mathrm{tot}}(t)+I(t)Q\rho _{%
\mathrm{tot}}(0),  \label{tcl2} \\
K(t) &=&-iPL\theta (t),\quad  \notag \\
I(t) &=&K(t)\exp (-iQLt),  \notag \\
\theta (t) &=&\left( P+\exp (-iQLt)Q\exp (iLt)\right) ^{-1},  \notag
\end{eqnarray}%
where $Q=I-P$. For simplicity, we restricted ourselves to the case where the
Liouvillian is time-independent. This is an exact, inhomogeneous, first
order linear differential equation. Both the TCL generator $K(t)$ of the
linear part and the inhomogeneity $I(t)$ are explicitly time-dependent
superoperators which do not depend on which initial state $\rho _{\mathrm{tot%
}}(0)$ is taken$.$ Observe that the inhomogeneous term vanishes if the
initial state satisfies the relation $Q\rho _{\mathrm{tot}}(0)=0$, i.e., if%
\begin{equation}
P\rho _{\mathrm{tot}}(0)=\rho _{\mathrm{tot}}(0).  \label{tcl3}
\end{equation}%
This coincides with having factorized initial conditions.\cite{Breuer:book}
Below, we consider when this is possible in our setting.

The condition that $P\rho _{\mathrm{tot}}$ contains all the relevant
information means that the following condition%
\begin{equation}
\rho =\mathrm{Tr}_{\vec{r}}\mathrm{\,}\rho _{\mathrm{tot}}=\mathrm{Tr}_{\vec{%
r}}\mathrm{\,}P\rho _{\mathrm{tot}}  \label{tcl4}
\end{equation}%
together with $P^{2}=P$ should be imposed on $P$ as a projection operator.
Let us take any orthogonal decomposition of the unit operator $I_{\vec{r}}$
on the coordinate state space $\mathrm{span\,}\left\{ f_{si},f_{t}\right\}
_{i=1,2,3}$%
\begin{equation}
I_{\vec{r}}=\sum_{i=1}^{3}\left\vert f_{si}\right\rangle \left\langle
f_{si}\right\vert +\left\vert f_{t}\right\rangle \left\langle
f_{t}\right\vert ,  \label{tcl5}
\end{equation}%
i.e., a collection of projection operators $\Pi _{a}$ that satisfy%
\begin{equation}
\Pi _{a}\Pi _{b}=\delta _{ab}\Pi _{b},\quad \sum_{a}\Pi _{a}=I_{\vec{r}}.
\label{tcl6}
\end{equation}%
Then, following Ref.~\onlinecite{Breuer:book} we can define a linear map by means of%
\begin{equation}
P\rho _{\mathrm{tot}}=\sum_{a}\mathrm{Tr}_{\vec{r}}\,\left\{ \Pi _{a}\rho _{%
\mathrm{tot}}\right\} \otimes \frac{1}{N_{a}}\Pi _{a},  \label{tcl7}
\end{equation}%
where $N_{a}=\mathrm{Tr}_{\vec{r}}\,\left\{ \Pi _{a}\right\} $. It is easy
to check that this superoperator fulfills the above conditions. As an
example, let us consider the case where $\Pi _{a}=\Pi
_{s}=\sum_{i=1}^{3}\left\vert f_{si}\right\rangle \left\langle
f_{si}\right\vert $ and $\Pi _{b}=\Pi _{t}=\left\vert f_{t}\right\rangle
\left\langle f_{t}\right\vert $. Then, the six-by-six density matrix%
\begin{equation}
\rho _{\mathrm{tot}}=\left( 
\begin{array}{cc}
\rho ^{ss} & \rho ^{st} \\ 
\rho ^{ts} & \rho ^{tt}%
\end{array}%
\right)  \label{tcl8}
\end{equation}%
is transformed under the map Eq.~(\ref{tcl7}) to%
\begin{equation}
P\rho _{\mathrm{tot}}=\left( 
\begin{array}{cc}
\rho _{\mathrm{av}}^{ss}I & 0 \\ 
0 & \rho ^{tt}%
\end{array}%
\right)  \label{tcl9}
\end{equation}%
where $\rho _{\mathrm{av}}^{ss}=\frac{1}{3}\sum_{i=1}^{3}\rho _{ii}^{ss}$.
Notice, however, that the projection superoperator $P$ must be realizable as
a superoperator independent of $\rho _{\mathrm{tot}}$, i.e., the matrix
state Eq.~(\ref{tcl9}) must be obtained by multiplying the matrix (\ref{tcl8}%
) from the left and right by some matrices independent of $\rho _{\mathrm{tot%
}}$. It is clear that by taking $\Pi _{a}=\left\{ \Pi _{si}=\left\vert
f_{si}\right\rangle \left\langle f_{si}\right\vert ,i=1,2,3;\,\Pi
_{t}\right\} $ the projection superoperator $P$ can be written as%
\begin{equation}
P\rho _{\mathrm{tot}}=\left( 
\begin{array}{cc}
\mathrm{diag}\,(\rho _{ii}^{ss}) & 0 \\ 
0 & \rho ^{tt}%
\end{array}%
\right) =\sum_{i=1}^{3}E_{si}\rho _{\mathrm{tot}}E_{si}+E_{t}\rho _{\mathrm{%
tot}}E_{t}  \label{tcl10}
\end{equation}%
where $\mathrm{diag}\,(\rho _{ii}^{ss})$ is a diagonal matrix with $\rho
_{ii}^{ss},$ $i=1,2,3$ entries on the diagonal; $E_{si}$ and $E_{t}$ are
orthogonal projection matrices independent of $\rho _{\mathrm{tot}}$ and
defined by%
\begin{equation}
E_{si}=\left( 
\begin{array}{cc}
1_{i} & 0 \\ 
0 & 0%
\end{array}%
\right) ,\quad E_{t}=\left( 
\begin{array}{cc}
0 & 0 \\ 
0 & I%
\end{array}%
\right) .  \label{tcl11}
\end{equation}%
Here, $1_{i}$ is a diagonal matrix with all zero entries except $1$ in the $%
i $-th entry.\ However, without changing matrix dimension it is not possible
to change the diagonal entries $\rho _{ii}^{ss}$ in Eq.(\ref{tcl10}) to $%
\rho _{\mathrm{av}}^{ss}$ by multiplying the matrix Eq.~(\ref{tcl10}) by
some diagonal matrices independent of $\rho _{\mathrm{tot}}$. Hence, in
order to comply with the requirement of\ state\ independency one should
extend the definition of the projection superoperator to Eq.~(\ref{tcl10}).
Observe that the state Eq.~(\ref{tcl10}) contains two extra pieces of
information:\ besides the probability of single state occupation, $\rho
_{11}^{ss}$, also, the probabilities of double state occupancy, $\rho
_{22}^{ss}$ and $\rho _{33}^{ss}$, which will be compressed after averaging $%
\mathrm{Tr}_{\vec{r}}\mathrm{\,}P\rho _{\mathrm{tot}}$ to $\rho _{\mathrm{av}%
}^{ss}$. Then, it follows from Eq.~(\ref{tcl10}) that%
\begin{equation}
Q\rho _{\mathrm{tot}}=E_{s}\rho _{\mathrm{tot}}E_{t}+E_{t}\rho _{\mathrm{tot}%
}E_{s}+\sum_{i\neq j}E_{si}\rho _{\mathrm{tot}}E_{sj}  \label{tcl12}
\end{equation}%
where $E_{s}=\sum_{i}E_{si}$.

In principle, one can \textquotedblleft compress\textquotedblright\ $\mathrm{%
diag}\,(\rho _{ii}^{ss})$ in Eq.~(\ref{tcl10}) by applying a rectangular
6-by-4 matrix%
\begin{equation}
C=\left( 
\begin{array}{cc}
\begin{array}{c}
1 \\ 
1 \\ 
1%
\end{array}
& 0_{3\times 3} \\ 
0_{3\times 1} & I_{3\times 3}%
\end{array}%
\right)  \label{tcl12.1}
\end{equation}%
so that%
\begin{eqnarray}
P\rho _{\mathrm{tot}} \longmapsto \rho &=& C^{T}\left(
\sum_{i=1}^{3}E_{si}\rho _{\mathrm{tot}}E_{si}+E_{t}\rho _{\mathrm{tot}%
}E_{t}\right) C  \notag \\
&=&\left( 
\begin{array}{cc}
\sum_{i}\rho _{ii}^{ss} & 0_{1\times 3} \\ 
0_{3\times 1} & \rho ^{tt}%
\end{array}%
\right) .  \label{tcl12.2}
\end{eqnarray}

However, the transformation Eq.~(\ref{tcl12.2}) is not a projection
operation: $C^{T}C\neq C$.

In numerical computations it is often convenient to rewrite matrix operator
equations in tensor product form (see for example Ref.~\onlinecite{GL96} and
references therein)%
\begin{equation}
Y=CXB^{T}\quad \Longleftrightarrow \quad \left\vert Y\right\rangle =\left(
B\otimes C\right) \left\vert X\right\rangle  \label{tcl13}
\end{equation}%
where $Y,C,X,B\in 
\mathbb{C}
^{n\times n}$, $B\otimes C\in 
\mathbb{C}
^{n^{2}\times n^{2}}$ are matrices and we introduced the Liouville vector
space $\left\vert X\right\rangle \in 
\mathbb{C}
^{n^{2}}$ by applying the \textquotedblleft vectorization\textquotedblright\
operation to a matrix:%
\begin{equation}
X\quad \longmapsto \quad \left\vert X\right\rangle =\left( 
\begin{array}{c}
X_{\cdot 1} \\ 
\vdots \\ 
X_{\cdot n}%
\end{array}%
\right) ,  \label{tcl14}
\end{equation}%
which amounts to a \textquotedblleft stacking\textquotedblright\ of matrix
columns. It is easy to check that $\left\langle X\right. \left\vert
Y\right\rangle =\mathrm{Tr}\,(X^{\dagger }Y)$.

Using these rules, Eqs. (\ref{tcl13}) and (\ref{tcl14}), we can rewrite the
superoperators in Eq.~(\ref{tcl2}) in the tensor product form%
\begin{equation}
\begin{array}{c}
L=I\otimes H-H^{T}\otimes I,\quad P=\sum\limits_{i=1}^{3}E_{si}\otimes
E_{si}+E_{t}\otimes E_{t}, \\ 
Q=E_{s}\otimes E_{t}+E_{t}\otimes E_{s}+\sum\limits_{i\neq
j}^{{}}E_{si}\otimes E_{sj}.%
\end{array}
\label{tcl15}
\end{equation}%
Then, with the use of identities {}$(A\otimes B)(C\otimes D)=AC\otimes BD$
and $(A+B)\otimes C=A\otimes C+B\otimes C$, one obtains

\begin{eqnarray}
QLQ &=&\sum_{ij}^{{}}\left( E_{t}\otimes
E_{si}HE_{sj}-E_{si}H^{T}E_{sj}\otimes E_{t}\right) +  \notag \\
&&(E_{s}\otimes E_{t}HE_{t}-E_{t}H^{T}E_{t}\otimes E_{s})+  \notag \\
&&\sum_{i}^{{}}\sum_{j\neq i,k\neq i}\left( E_{si}\otimes
E_{sj}HE_{sk}\right. -  \notag \\
&&\left. E_{sj}H^{T}E_{sk}\otimes E_{si}\right) +  \notag \\
&&\sum_{i}^{{}}\sum_{j\neq i}\left\{ E_{si}\otimes \left(
E_{t}HE_{sj}+E_{sj}HE_{t}\right) \right. -  \notag \\
&&\left. \left( E_{t}H^{T}E_{sj}+E_{sj}H^{T}E_{t}\right) \otimes
E_{si}\right\} ,  \label{tcl16}
\end{eqnarray}

\begin{eqnarray}
PLP &=&\sum_{i}\left( E_{si}\otimes E_{si}HE_{si}-E_{si}H^{T}E_{si}\otimes
E_{si}\right) +  \notag \\
&&E_{t}\otimes E_{t}HE_{t}-E_{t}H^{T}E_{t}\otimes E_{t},  \label{tcl17}
\end{eqnarray}

\begin{eqnarray}
QLP &=&E_{t}\otimes E_{s}HE_{t}-E_{s}H^{T}E_{t}\otimes E_{t}+  \notag \\
&&\sum_{i}^{{}}\left( E_{si}\otimes E_{t}HE_{si}-E_{t}H^{T}E_{si}\otimes
E_{si}\right) +  \notag \\
&&\sum_{i\neq j}^{{}}\left( E_{si}\otimes
E_{sj}HE_{si}-E_{si}H^{T}E_{sj}\otimes E_{sj}\right) ,  \notag \\
PLQ &=&(QLP)^{\dagger }.  \label{tcl18}
\end{eqnarray}

Having obtained the basic superoperators in tensor product form, in the next
step we can consider how to compute the superoperators $K(t)$ and $I(t)$ in
Eq.~(\ref{tcl2}) and then find the conditions under which
Eq.~(\ref{tcl3}) (factorized initial conditions)
is fulfilled. Inspection of Eq.~(\ref{tcl2}) shows that in order to
calculate $K(t)$ and $I(t)$ one needs to calculate the exponentials $\exp
(-iQLQt)$ and $\exp (iLt)$ and take an inverse of a matrix to obtain $\theta
(t)$. Calculation of $\exp (iLt)$ does not introduce any difficulties and
can be reduced to the eigenvalue problem Eq.~(\ref{Eq37}), namely, the
eigenvalue problem for the Liouvillean superoperator%
\begin{eqnarray}
\left( I\otimes H-H^{T}\otimes I\right) \left\vert E_{nm}\right\rangle
&=&\varepsilon _{nm}\left\vert E_{nm}\right\rangle ,\quad  \label{tcl19} \\
n,m &=&1,2,\cdots ,6  \notag
\end{eqnarray}%
is solved in terms of eigenvalue solutions to Eq.~(\ref{Eq37}): $\varepsilon
_{nm}=\varepsilon _{n}-\varepsilon _{m}$ and $E_{nm}=\left\vert
e_{n}\right\rangle \left\langle e_{m}\right\vert $ where%
\begin{equation*}
\left\vert e_{n}\right\rangle =\left( 
\begin{array}{c}
\left\vert e_{sn}\right\rangle \\ 
\left\vert e_{tn}\right\rangle%
\end{array}%
\right)
\end{equation*}%
and $\left\vert E_{nm}\right\rangle $ is obtained according to the rule Eq.~(%
\ref{tcl14}). Then,%
\begin{equation}
\exp (iLt)=\sum_{nm}\exp (i\varepsilon _{nm}t)\left\vert E_{nm}\right\rangle
\left\langle E_{nm}\right\vert .  \label{tcl20}
\end{equation}

A direct calculation of $\exp (-iQLQt)$ does not seem to have a similar
simple solution due to the rather complicated tensor structure of $QLQ$ Eq.~(%
\ref{tcl16}). However, one can overcome this problem by simplifying $\theta
(t)$ to%
\begin{equation}
\begin{array}{lll}
\theta (t) & = & \left( \exp (-iQLQt)\left[ \exp (iQLQt)P+Q\exp (iLt)\right]
\right) ^{-1} \\ 
& = & \left( P+Q\exp (iLt)\right) ^{-1}\exp (iQLQt),%
\end{array}
\label{tcl21}
\end{equation}%
where we have used the identity $\exp (iQLQt)P=P$ and unitarity of $\exp
(-iQLQt)$. Since the $\theta (t)$ superoperator in $K(t)$ operates on the
state $P\left\vert \rho _{\mathrm{tot}}\right\rangle $ and the $Q$%
-projected exponential is cancelled in the $I(t)$ superoperator, the TCL
equation can be rewritten in the form%
\begin{eqnarray}
\frac{d}{dt}\left\vert P\rho _{\mathrm{tot}}(t)\right\rangle &=&-iK(t)\left(
\left\vert P\rho _{\mathrm{tot}}(t)\right\rangle +\left\vert Q\rho _{\mathrm{%
tot}}(0)\right\rangle \right)  \label{tcl22} \\
K(t) &=&PL\left( P+Q\exp (iLt)\right) ^{-1}  \label{tcl23}
\end{eqnarray}%
free from $\exp (-iQLQt)$. In fact, as one can see $K(t)$ and $I(t)$ are the
same superoperators.

Eq.~(\ref{tcl22}) along with (\ref{Eq36}) are our central results. Knowing
how to calculate the exponential Eq.~(\ref{tcl20}), it is then
straightforward to compute $K(t)$ Eq.~(\ref{tcl23}) by applying standard
matrix multiplication and inversion routines.

\subsection{Separation of unitary and non-unitary dynamics}

Condition (\ref{tcl3}) for a factorized initial state can be rewritten as%
\begin{equation}
Q\rho _{\mathrm{tot}}(0)=\left( 
\begin{array}{cc}
\rho ^{ss}(0)-\mathrm{diag}\,(\rho _{ii}^{ss}(0)) & \rho ^{st}(0) \\ 
\rho ^{ts}(0) & 0%
\end{array}%
\right) =0.  \label{tcl24}
\end{equation}%
It can be fulfilled exactly in the following two cases: (i) the singlet
case: $a_{t}(0)=0$ and any of the singlet state amplitudes $a_{si}(0)\neq 0$
with all the others being zero, $a_{sj}(0)=0$, $j\neq i$; (ii) the triplet
case: $a_{s}(0)=0$. In other cases, including the mixed one, Eq.~(\ref{tcl22}%
) will not be in closed form and will contain a non-zero inhomogeneous term.
Notice that unlike the Lindblad formulation [Eq.~(\ref{Eq36})], the superoperator $K(t)$ Eq.~(\ref{tcl23}%
) does not depend on which initial state -- singlet or triplet -- is taken.

Henceforth we neglect the inhomogeneity for simplicity. To separate
unitary effects from non-unitary ones, we decompose the dynamics
generator $K(t)$ into Hermitian and non-Hermitian parts:
\begin{eqnarray}
K(t) &=&K_{+}(t)+K_{-}(t),  \notag \\
K_{\pm }(t) &=&\frac{1}{2}\left\{ K(t)\pm K^{\dagger }(t)\right\} .
\label{tcl25}
\end{eqnarray}%
Transforming to the interaction representation (denoted by a superscript
hat) we have for the unitarily transformed state%
\begin{equation}
\left\vert P\hat{\rho}_{\mathrm{tot}}(t)\right\rangle =T_{\rightarrow }\exp
\left( i\int_{0}^{t}K_{+}(\tau )\,d\tau \right) \left\vert P\rho _{\mathrm{%
tot}}(t)\right\rangle  \label{tcl26}
\end{equation}%
the following equation of motion:%
\begin{equation}
\frac{d}{dt}\left\vert P\hat{\rho}_{\mathrm{tot}}(t)\right\rangle =-i\hat{K}%
_{-}(t)\left\vert P\hat{\rho}_{\mathrm{tot}}(t)\right\rangle ,  \label{tcl27}
\end{equation}%
where%
\begin{eqnarray}
\hat{K}_{-}(t) &=&T_{\rightarrow }\exp \left( i\int_{0}^{t}K_{+}(\tau
)\,d\tau \right) \times  \notag \\
&&K_{-}(t)T_{\leftarrow }\exp \left( -i\int_{0}^{t}K_{+}(\tau )\,d\tau
\right) .  \label{tcl28}
\end{eqnarray}%
Here, $(T_{\rightarrow })T_{\leftarrow }$ denotes the (anti)-chronological
ordering operator. One can see that Eq.\ (\ref{tcl27}) describes non-unitary
dynamics since the generator $\hat{K}_{-}^{\dagger }(t)=-\hat{K}_{-}(t)$ is
an anti-Hermitian superoperator. If $K_{-}(t)\equiv 0$, then the evolution
is unitary%
\begin{equation}
\left\vert P\rho _{\mathrm{tot}}(t)\right\rangle =T_{\leftarrow }\exp \left(
-i\int_{0}^{t}K_{+}(\tau )\,d\tau \right) \left\vert P\rho _{\mathrm{tot}%
}(0)\right\rangle .  \label{tcl29}
\end{equation}

\subsection{Short time expansion}

Expanding $K(t)$ in powers of $t$, one obtains%
\begin{eqnarray}
K_{+}(t) &=&PLP+\left( \left\{ (QLP)^{\dagger }(QLP),PLP\right\} \right. - 
\notag \\
&&\left. 2(QLP)^{\dagger }(QLQ)(QLP)\right) \frac{t^{2}}{4}+\cdots ,  \notag
\\
K_{-}(t) &=&-i(QLP)^{\dagger }(QLP)t+  \notag \\
&&\left[ (QLP)^{\dagger }(QLP),PLP\right] \frac{t^{2}}{4}+\cdots
\label{tcl30}
\end{eqnarray}%
where $\left\{ \cdot ,\cdot \right\} $ denotes the anti-commutator. In the
zeroth order approximation $K(t)\equiv PLP$, Eq.~(\ref{tcl22}) is reduced to
a unitary matrix equation of the form%
\begin{equation}
\left\{ 
\begin{array}{l}
\dot{\rho}_{ii}^{ss}=0,\quad i=1,2,3 \\ 
\dot{\rho}^{tt}=-i[H^{tt},\rho ^{tt}]%
\end{array}%
\right. ,  \label{tcl31}
\end{equation}%
which describes a unitary evolution of the triplet state density matrix $%
\rho ^{tt}$ under the Hamiltonian $H^{tt}$ with the singlet state population
probabilities$\ \rho _{ii}^{ss}(t)=\rho _{ii}^{ss}(0)$ being constant in
time. Let us consider the non-unitary effects which are induced by the first
term proportional to $t$ in the expansion of $K_{-}(t)$ Eq.~(\ref{tcl30}).

\subsection{Purity}

Observe that $\mathrm{Tr}\,P\rho _{\mathrm{tot}}=\mathrm{Tr}\,\rho $ but $%
\mathrm{Tr}\,(P\rho _{\mathrm{tot}})^{2}\neq \mathrm{Tr}\,\rho ^{2}$. In
order to obtain $\rho $ from $P\rho _{\mathrm{tot}}$ we first have to apply the
``compression'' transformation Eq.~(\ref{tcl12.2}) so that the purity becomes%
\begin{equation}
p(t)=\mathrm{Tr}\,\rho ^{2}(t)=\left\langle P\rho _{\mathrm{tot}%
}(t)\right\vert \left. CC^{T}P\rho _{\mathrm{tot}}(t)CC^{T}\right\rangle
\label{tcl32}
\end{equation}%
where%
\begin{equation}
CC^{T}=\left( 
\begin{array}{cc}
J_{3\times 3} & 0 \\ 
0 & I_{3\times 3}%
\end{array}%
\right) ,\quad J_{3\times 3}=\left( 
\begin{array}{ccc}
1 & 1 & 1 \\ 
1 & 1 & 1 \\ 
1 & 1 & 1%
\end{array}%
\right)  \label{tcl33}
\end{equation}%
Then, using Eq.~(\ref{tcl22}) we obtain an expression for the time derivative%
\begin{eqnarray}
\frac{d}{dt}\mathrm{Tr}\,\rho ^{2}(t) &=&i\left\langle P\rho _{\mathrm{tot}%
}(t)\right\vert (K^{\dagger }(t)CC^{T}-  \notag \\
&&\left. CC^{T}K(t))P\rho _{\mathrm{tot}}(t)CC^{T}\right\rangle .
\label{tcl34}
\end{eqnarray}

It is easy to check that the first term, $PLP$, in the expansion of $K(t)$
does not contribute to Eq.~(\ref{tcl34}) and the first non-zero contribution
comes from the first term in the expansion of $K_{-}(t)$ [Eq.~(\ref{tcl30}%
)], which can be written as%
\begin{equation}
\frac{d}{dt}\mathrm{Tr}\,\rho ^{2}(t)=-2t\left\langle QLP\rho _{\mathrm{tot}%
}(0)\right\vert \left. QLPCC^{T}P\rho _{\mathrm{tot}}(0)CC^{T}\right\rangle .
\label{tcl35}
\end{equation}%
Further, with the use of Eq.~(\ref{tcl18}) one obtains%
\begin{eqnarray}
QLP\rho _{\mathrm{tot}}(0) &=&\left( 
\begin{array}{cc}
c_{11} & c_{12} \\ 
-c_{12}^{\dagger } & 0%
\end{array}%
\right)  \label{tcl36} \\
c_{11} &=&[H^{ss},\mathrm{diag\,}(\rho ^{ss}(0))],\quad  \notag \\
c_{12} &=&H^{st}\rho ^{tt}(0)-\mathrm{diag\,}(\rho ^{ss}(0))H^{st},  \notag
\end{eqnarray}

and%
\begin{eqnarray}
QLPCC^{T}P\rho _{\mathrm{tot}}(0)CC^{T} &=&\left( 
\begin{array}{cc}
0 & d_{12} \\ 
d_{21} & 0%
\end{array}%
\right) ,  \label{tcl37} \\
d_{12} &=&H^{st}\rho ^{tt}(0)-\rho _{a}^{ss}(0)H^{st},  \notag \\
d_{21} &=&H^{ts}\rho _{a}^{ss}(0)-\rho ^{tt}(0)H^{ts},  \notag
\end{eqnarray}%
where $\rho _{a}^{ss}(0)=\sum_{i}\rho _{ii}^{ss}(0)J_{3\times 3}$ and Eq.~(%
\ref{tcl35}) is reduced to%
\begin{eqnarray}
\frac{d}{dt}\mathrm{Tr}\,\rho ^{2}(t) &=&-2t\left\{ \mathrm{Tr}\,(H^{ts}\rho
_{a}^{ss}(0)-\rho ^{tt}(0)H^{ts})\right. \times  \notag \\
&&(\mathrm{diag\,}(\rho ^{ss}(0))H^{st}-H^{st}\rho ^{tt}(0))+  \notag \\
&&\mathrm{Tr}\,(H^{st}\rho ^{tt}(0)-\rho _{a}^{ss}(0)H^{st})\times  \notag \\
&&\left. (\rho ^{tt}(0)H^{ts}-H^{ts}\mathrm{diag\,}(\rho ^{ss}(0)))\right\}
\label{tcl38}
\end{eqnarray}

Finally, from Eq.~(\ref{tcl38}) one obtains%
\begin{equation}
\frac{d}{dt}\mathrm{Tr}\,\rho ^{2}(t)=-4t\left\{ 
\begin{array}{l}
(H^{st}H^{ts})_{ii}, \\ 
\text{\textrm{singlet case: }}\rho _{ii}^{ss}(0)\neq 0; \\ 
\left\langle a_{t}(0)\right\vert H^{ts}H^{st}\left\vert
a_{t}(0)\right\rangle , \\ 
\text{$\mathrm{tr}$\textrm{iplet case: }}\rho ^{tt}(0)\neq 0,%
\end{array}%
\right.  \label{tcl39}
\end{equation}%
which is equivalent to the results we obtained in the Lindblad
analysis for the purity, Eqs.~(\ref{Eq57}) and (\ref{Eq49-1}) respectively. We provide a detailed discussion of the behavior of the
purity and other physically relevant quantities in Part II.\cite{KLII}

\subsection{TCL vs Lindblad-type dynamics}

Note that the Lindblad-type and TCL Eqs. (\ref{Eq36}) and (\ref{tcl22}) are
not mathematically equivalent formulations. In the Lindblad-type formulation
we allowed correlations between initial singlet and triplet states to be
incorporated into the dynamics, whereas in the TCL formulation we assumed
that all operators appearing in the dynamical equations are totally state
independent. The first approach gave us the flexibility needed to cover
mixed states, whereas the TCL equations become non-closed [inhomogeneous
term in (\ref{tcl22}) is non-zero]\ in the mixed state case.

\section{Summary, Discussion, and Conclusions}

\label{sec:concI}

In this work we have shown that the time-evolution of the spin-density
matrix $\rho (t)$ that describes a system of two electron spins is in
general non-unitary when spatial degrees of freedom are accounted for,
without coupling to a \textquotedblleft true\textquotedblright\ bath. The
non-unitary effects (e.g., the total spin is unconserved) in the pure-spin
evolution are due to a non-zero coupling between singlet and triplet states.
Our primary focus in this work was the derivation of dynamical equations for
the spin-density matrix, and a corrsponding study of the non-unitary
effects. Invoking standard ideas from the theory of open quantum systems,
one can define a \textquotedblleft system + bath\textquotedblright\ by
formally associating the spin variables with the \textquotedblleft
system\textquotedblright\ degrees of freedom, and the spatial variables with
the \textquotedblleft bath\textquotedblright\ degrees of freedom. With
system and bath thus defined, one can apply the standard machinery of open
quantum systems theory to attack the above problem. We did this by first
constructing an analog of the Kraus operator sum representation for $\rho
(t) $ and then deriving master equations for $\rho (t)$ in the Lindblad and
TCL forms.

The Kraus representation [Eq.~(\ref{Eq14})]\ has the shortcoming that its
Kraus operators are dependent on the initial state $\rho (0)$. This is a
consequence of the fact that an arbitrary initial state $\rho (0)$ cannot be
represented as a product state with some reference state in coordinate space
due to the Pauli antisymmetry principle, which is imposed on the total
wavefunction. As a result of this dependence, the complete positivity
property of the mapping (\ref{Eq14}) is not guaranteed.

The Lindblad-type master equation Eq.~(\ref{Eq36}) describes the \emph{exact}
dynamics of $\rho (t)$ (it is not a Markovian approximation in our case).
Again, the generators of this dynamics are not totally independent of
initial conditions. This is to be expected; after all, the $\rho $-dynamics
is inherited from the total system unitary dynamics described by Eq.~(\ref%
{Eq10}). In the total Hilbert space, the state is defined by $11$ real
parameters [$6$ complex amplitudes $\{a_{s}(t),a_{t}(t)\}$ minus the
normalization condition], while $\rho $ is defined by $5$ real parameters
(\textquotedblleft system degrees of freedom\textquotedblright ): $3$
amplitude moduli $|a_{ti}(t)|$, $i=1,2,3$, and two relative phase angles
between the amplitudes $a_{t1}(t)$, $a_{t2}(t)$ and $a_{t1}(t)$, $a_{t3}(t)$
[$\rho _{tij}(t)=a_{ti}(t)a_{ti}^{\ast }(t)$, $\rho _{s}(t)=1-\sum_{i}\rho
_{tii}(t)$]. Thus, we have $6$ extra degrees of freedom [$4$ real parameters
defining two complex double-ocupancy amplitudes, $a_{s2}(t)$ and $a_{s3}(t)$%
, and two phases of the amplitudes $a_{s1}(t)$ and $a_{t1}(t)$]\ in the
reduction $\Psi _{\mathrm{tot}}\rightarrow \rho $, which should be present
in the dynamics equation if it is \emph{exact}. The three complex-valued
equations (\ref{Eq29m}), that establish relationships between amplitudes $%
a_{s}$ and $a_{t}$ at the initial moment, allow us to express $6$ extra
degrees of freedom as a function of system degrees of freedom. Fixing the
correlation matrix $R_{m}(0)$ appearing there, defines a domain in the space
of $11$ real parameters so that dynamical operators in Eq.~(\ref{Eq36})
acquire a dependence on the domain on which a correlated mixed state is
defined. Of course, during time evolution $R_{m}(t)$ evolves according to
Eq.~(\ref{Eq30m+1}) and the $\{a_{s}(t),a_{t}(t)\}$ point can move out of
the domain, fixed by $R_{m}(0)$, on which the initial states were defined.
In this case, $R_{m}(t)$ will define a new domain at time $t$, which can be
considered as the initial state domain for later times.

It turns out that the Heisenberg interaction does not affect the
pure-spin dynamics as long as the coupling between singlet and triplet
states is 
neglected -- see Eq.~(\ref{Eq28-3}). If there is a non-zero coupling, the
Heisenberg interaction modifies the corresponding non-unitary terms: the
Heisenberg interaction constant $J_{H}$ appears in Eq.~(\ref{Eq44}). It also
appears in the unitary Lamb shift term [the matrix $P_{t}$, which defines this
shift, depends on $J_{H}$]. Note that the matrix functions in Eq.~(\ref{Eq44}%
), responsible for the singlet-triplet coupling, depend quadratically on the
magnitude of the  interaction between singlet and triplet states, described by the $%
H^{ts} $ matrix elements.

The TCL Eq.~(\ref{tcl22}) is also \emph{exact} and describes a non-unitary
evolution; the dynamical generator $K(t)$ is non-Hermitian and does not
depend on initial conditions. However, the TCL equation is not closed with
respect to the $\left\vert P\rho _{\mathrm{tot}}(t)\right\rangle $ state
since it contains an inhomogeneous term, which is non-zero in the mixed
initial state case. In the non-mixed case defined by Eq.~(\ref{tcl24}), when
the TCL equation becomes closed, we found a short-time expansion [Eq.~(\ref%
{tcl39})] for the purity function $p(t)$, whose time-dependence signifies
non-unitary effects in time-evolution. For consistency, we obtained the same
short-time formulas within the Lindblad-type formulation, Eqs. (\ref{Eq49-1}%
) and (\ref{Eq57}). In Part II,\cite{KLII} we provide a detailed numerical
example demonstrating these non-unitary effects.

In conclusion, the spin-density matrix completely describes the pure-spin
dynamics of a two-electron system in the case where the electrons cannot be
spatially resolved. The formalism we have developed can be generalized in
several directions. First, one could take into consideration excited-state
orbitals, which will increase the dimensionality of the total Hilbert space.
Another interesting generalization is to consider several interacting QDs
along the lines developed in Refs.~\onlinecite{MizelLidar:04,MizelLidar:04a,WoodworthMizelLidar:05} and to derive for this
few-body case the corresponding spin-density matrix dynamics.

\begin{acknowledgments}
This work was supported by NSF Grant No.\ CCF-0523675.
\end{acknowledgments}


\end{document}